\documentclass[10pt,journal,compsoc]{IEEEtran}
\usepackage{graphicx}
\usepackage[lowtilde]{url}
\usepackage[caption=false,font=footnotesize]{subfig}
\usepackage{pgfplots}
\pgfplotsset{compat=newest}
\usetikzlibrary{plotmarks}
\usepackage{grffile}
\usepackage{amsmath}
\usepackage{multirow}
\usepackage{enumitem}
\usepackage[euler]{textgreek}
\usepackage{booktabs} 
\usepackage{array}
\usepackage{verbatim}
\usepackage{ragged2e}
 \usepackage{cite}

\makeatletter
\long\def\@makecaption#1#2{\ifx\@captype\@IEEEtablestring%
\footnotesize\begin{center}{\normalfont\footnotesize #1}\\
{\normalfont\footnotesize\scshape #2}\end{center}%
\@IEEEtablecaptionsepspace
\else
\@IEEEfigurecaptionsepspace
\setbox\@tempboxa\hbox{\normalfont\footnotesize {#1.}~~ #2}%
\ifdim \wd\@tempboxa >\hsize%
\setbox\@tempboxa\hbox{\normalfont\footnotesize {#1.}~~ }%
\parbox[t]{\hsize}{\normalfont\footnotesize \noindent\unhbox\@tempboxa#2}%
\else
\hbox to\hsize{\normalfont\footnotesize\hfil\box\@tempboxa\hfil}\fi\fi}
\makeatother

\newcommand{\RN}[1]{%
  \textup{\uppercase\expandafter{\romannumeral#1}}%
}

\newcolumntype{C}[1]{>{\centering\let\newline\\\arraybackslash\hspace{0pt}}m{#1}}

\renewcommand{\justify}{\leftskip=0pt \rightskip=0pt plus 0cm}

\begin{document}

\title{Perceptual Quality Assessment of Colored 3D Point Clouds}

\author{Honglei~Su,~\IEEEmembership{Member,~IEEE,}~Qi~Liu, ~Zhengfang~Duanmu,~\IEEEmembership{Student Member,~IEEE,}~Wentao~Liu,~\IEEEmembership{Student Member,~IEEE,}~and~Zhou~Wang,~\IEEEmembership{Fellow,~IEEE}
\IEEEcompsocitemizethanks{\IEEEcompsocthanksitem Honglei Su and Qi Liu are with the College of Electronic Information, Qingdao University, Qingdao, 266071, China (e-mail: suhonglei@qdu.edu.cn, sdqi.liu@gmail.com).\protect\\
\IEEEcompsocthanksitem Zhengfang Duanmu, Wentao Liu and Zhou Wang are with the Department of Electrical and Computer Engineering, University of Waterloo, Waterloo, ON, N2L 3G1, Canada (e-mail: \{zduanmu, w238liu, zhou.wang\}@uwaterloo.ca).

}
\thanks{Corresponding author: Qi Liu.}}

\markboth{}%
{Shell \MakeLowercase{\textit{et al.}}: Bare Demo of IEEEtran.cls for Computer Society Journals}

\IEEEtitleabstractindextext{%
\begin{abstract}
\justify
The real-world applications of 3D point clouds have been growing rapidly in recent years, but not much effective work has been dedicated to perceptual quality assessment of colored 3D point clouds. In this work, we first build a large 3D point cloud database for subjective and objective quality assessment of point clouds. We construct 20 high quality, realistic, and omni-directional point clouds of diverse contents. We then apply downsampling, Gaussian noise, and three types of compression algorithms to create 740 distorted point clouds. We carry out a subjective experiment to evaluate the quality of distorted point clouds. Our statistical analysis finds that existing objective point cloud quality assessment (PCQA) models only achieve limited success in predicting subjective quality ratings. We propose a novel objective PCQA model based on the principle of information content weighted structural similarity. Our experimental results show that the proposed model well correlates with subjective opinions and significantly outperforms the existing PCQA models. The database has been made publicly available to facilitate reproducible research at https://github.com/qdushl/Waterloo-Point-Cloud-Database.
\end{abstract}

}

\maketitle

\IEEEdisplaynontitleabstractindextext

\IEEEpeerreviewmaketitle

\section{Introduction}\label{sec:intro}

\IEEEPARstart {A} 3D point cloud is a collection of points representing a 3D shape, object or environment. 
Each point has its own geometric coordinates and optional associated attributes such as color and surface normal. 
3D point clouds~\cite{ye2021meta,chen2019multi,zhang2020pointfilter,zhao2021relationship} find a wide variety of applications in manufacturing, construction, environmental monitoring, navigation, animation, etc. 
Many of these applications often require high quality point clouds that can truthfully reflect the geometry and perceptual attributes of the physical world.
However, various distortions may be introduced during the acquisition, compression~\cite{liu2020model}, transmission, storage, rendering processes, leading to a point cloud of degraded quality. 
Therefore, point cloud quality assessment (PCQA) draws a lot of attention from the research community~\cite{liu2021reduced,javaheri2017subjectiveA,alexiou2017subjective,alexiou2018point,alexious2018point,javaheri2017subjectiveB,torlig2018novel,zerman2019subjective,alexiou2017performance,alexiou2017towards,alexiou2018pointAS,alexiou2018impact,alexiou2018benchmarking,zhang2014subjective,nehme2019comparison,alexiou2019exploiting,da2019point,alexiou2019comprehensive,javaheri2020point,tian2017geometric,tian2017evaluation,tian2017updated,mekuria2017design,mekuria2017performance,meynet2019pc,dumic2018subjective,su2019perceptual,javaheri2020generalized,zerman2020textured,alexiou2020towards,alexiou2020pointxr,meynet2020pcqm,viola2020color,javaheri2020improving,javaheri2020mahalanobis,dumic2020point,viola2020reduced,cao2020visual,yang2020inferring,yang2020predicting,diniz2020multi,diniz2020towards,perry2020quality,hua2021cpc,diniz2021novel,diniz2020local,wu2021subjective,he2021towards,diniz2021color,hua2020vqa,liu2020point,hua2021bqe,tao2021point,xu2021epes,zhang2021ms}.

Since the human visual system (HVS) is the ultimate receiver in most applications, subjective quality assessment is the most straightforward and reliable approach to evaluate point cloud quality. 
A comprehensive subjective user study on a large-scale point cloud database brings several key benefits.
First, it provides an opportunity to study human behaviors in evaluating perceived quality of point clouds. 
Second, it offers a benchmark to validate and compare existing objective PCQA models.
Third, it supplies ground-truth data for objective PCQA model development.
Fourth, a high quality point cloud database can also serve as a playground for a variety of point cloud processing algorithms, such as denoising~\cite{zeng20193d} or compression~\cite{schwarz2018emerging}.
However, existing publicly available point cloud databases, such as the Point Cloud Library~\cite{rusu20113d}, the MPEG point cloud datasets~\cite{MPEG2017Datasets}, the JPEG Pleno database~\cite{JPEGPleno}, and the Stanford 3D scanning repository~\cite{turk1994zippered} often suffer from inferior acquisition quality, constrained viewpoints, and insufficient content types. 
As a consequence, subjective experiments derived from these databases~\cite{javaheri2017subjectiveA,alexiou2017subjective,alexiou2018point,alexious2018point,javaheri2017subjectiveB,torlig2018novel,zerman2019subjective,alexiou2017performance,alexiou2017towards,alexiou2018pointAS,alexiou2018impact,alexiou2018benchmarking,zhang2014subjective,nehme2019comparison,alexiou2019exploiting,da2019point,alexiou2019comprehensive,javaheri2020point,gutierrez2020quality,viola2020color,alexiou2020pointxr,zerman2020textured,javaheri2020improving,hua2020vqa,yang2020predicting,perry2020quality,liu2020point,nehme2020visual,wu2021subjective} are inherently deficient in serving the aforementioned four purposes.
\begin{figure*}[t]
  \centering
    \subfloat[]{\includegraphics[width=0.15\textwidth]{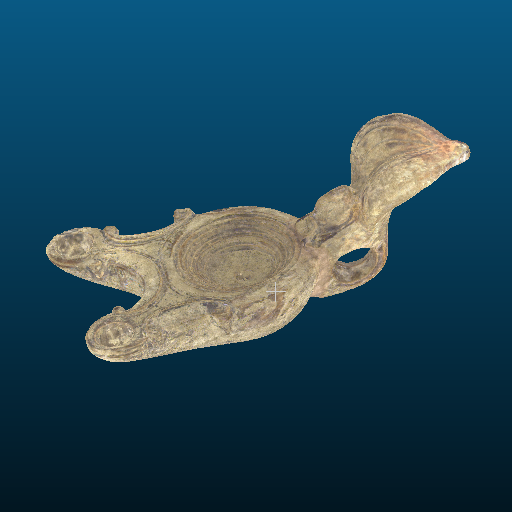}}\hskip0.1em
    \subfloat[]{\includegraphics[width=0.15\textwidth]{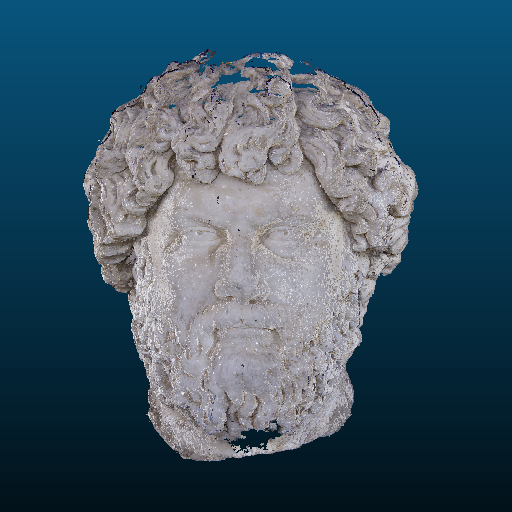}}\hskip0.1em
    \subfloat[]{\includegraphics[width=0.15\textwidth]{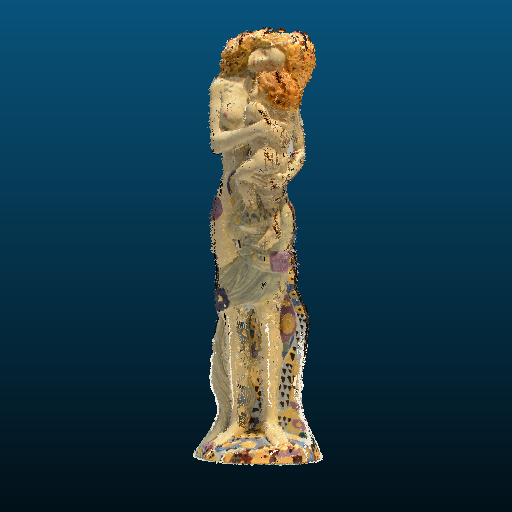}}\hskip0.1em
    \subfloat[]{\includegraphics[width=0.15\textwidth]{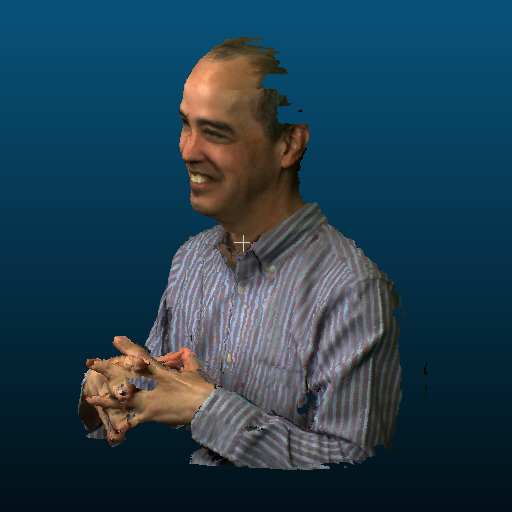}}\hskip0.1em
    \subfloat[]{\includegraphics[width=0.15\textwidth]{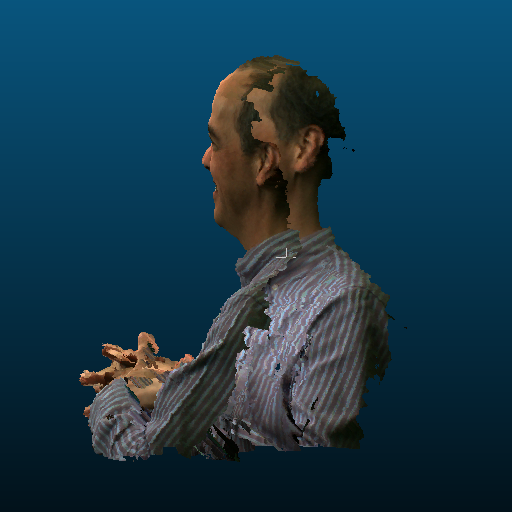}}\hskip0.1em
    \subfloat[]{\includegraphics[width=0.15\textwidth]{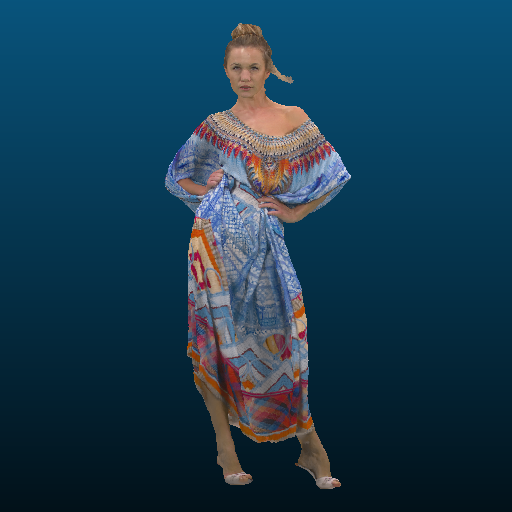}}\hskip0.1em
    \caption{Examples of point clouds from existing point cloud databases. (a) RomanOilLight. (b) Head. (c) Statue\_Klimt. (d) Phil. (e) Phil2. (f) Longdress.}
    \label{fig:otherpc}
\end{figure*}
On the other hand, objective PCQA models are highly desired for practical use, as a subjective test is often expensive, labor-intensive and time-consuming, restricting its usage in real-world applications.
Recently, substantial effort has been made to develop objective PCQA models~\cite{tian2017geometric,mekuria2017design,mekuria2017performance,alexiou2018pointAS,meynet2019pc,javaheri2020generalized,javaheri2020improving,javaheri2020mahalanobis,alexiou2020towards,meynet2020pcqm,viola2020color,diniz2020towards,diniz2020multi,hua2020vqa,hua2021cpc,yang2020inferring,torlig2018novel,alexiou2019exploiting,alexiou2019comprehensive,su2019perceptual,yang2020predicting,diniz2021novel,diniz2020local,wu2021subjective,he2021towards,diniz2021color,hua2020vqa,liu2020point,hua2021bqe,tao2021point,xu2021epes,zhang2021ms}.
Existing models either calculate distortions by directly comparing the 3D structures between the reference and distorted point clouds~\cite{tian2017geometric,mekuria2017design,mekuria2017performance,alexiou2018pointAS,meynet2019pc,javaheri2020generalized,javaheri2020improving,javaheri2020mahalanobis,alexiou2020towards,meynet2020pcqm,viola2020color,diniz2020towards,diniz2020multi,hua2020vqa,hua2021cpc,yang2020inferring,diniz2021novel,diniz2020local,diniz2021color,hua2020vqa,liu2020point,hua2021bqe,xu2021epes,zhang2021ms}, or leverage well-developed FR image quality assessment (IQA) models after projecting point clouds to multiple 2D images~\cite{torlig2018novel,alexiou2019exploiting,alexiou2019comprehensive,su2019perceptual,yang2020predicting,wu2021subjective,he2021towards,tao2021point}.
However, the performances of these models are often degraded by the absence of proper modelings of the HVS~\cite{wang2004image,wang2009mean} and/or the irregular data representation of point clouds~\cite{alexiou2019comprehensive,alexiou2019exploiting}.
Besides, none of these models is validated on large-scale subject-rated PCQA databases with diverse, high quality original point clouds, making the generalizability of existing PCQA models questionable. 

\begin{table*}[t]
\centering
\caption{Comparison of existing publicly available subject-rated PCQA databases}
\label{tab:publicly available}
\scalebox{0.8}{
  \begin{tabular}{c|c|c|c|c}
  \toprule
      Database name & Attribute & Source contents & Distortion type & Subject-rated point clouds\\\hline
      IRPC~\cite{javaheri2020point} & None, Color & 6 & PCL, G-PCC, V-PCC & 54\\
      vsenseVVDB~\cite{zerman2019subjective} & Color & 2 & V-PCC & 32\\
      vsenseVVDB2~\cite{zerman2020textured} & Color & 8 & Draco+JPEG, G-PCC, V-PCC & 164\\
      G-PCD~\cite{alexiou2017performance,alexiou2017towards} & None & 5 & Octree-puring, Gassian noise & 40\\
      RG-PCD~\cite{alexiou2018point} & None & 6 & Octree-puring & 24\\
      M-PCCD~\cite{alexiou2019comprehensive} & Color & 8 & G-PCC, V-PCC & 244\\
      PointXR~\cite{alexiou2020pointxr} & Color & 5 & G-PCC & 100\\
      NBU-PCD 1.0~\cite{hua2020vqa} & Color & 10 & Octree & 160\\
      CPCD 2.0~\cite{hua2021cpc} & Color & 10 & G-PCC, V-PCC, Gassian noise & 360\\
      SJTU-PCQA~\cite{yang2020predicting} & Color & 10 & Octree, downsampling, color and geometry noise & 420\\
      ICIP2020~\cite{perry2020quality} & Color & 6 & G-PCC, V-PCC & 90\\
      3DMDC~\cite{nehme2020visual} & Color & 5 & QGeo, QCol, SGeo, SCol & 80\\
      SIAT-PCQD~\cite{wu2021subjective} & Color & 20 & V-PCC & 340\\
      \textbf{WPC} & Color & \textbf{20} & Gassian noise, dowsampling, G-PCC, V-PCC & \textbf{740}\\
  \bottomrule
  \end{tabular}
}
\end{table*}

\begin{table*}[t]
\centering
\caption{Experiment settings of existing subjective tests for PCQA}
\label{tab:subjectivetest}
\scalebox{1.0}{
  \begin{tabular}{c|c|c|c|c}
  \toprule
      Literature & Methodology & Display device & Interaction method & Rendering mode\\\hline
      Javaheri \textit{et al.}~\cite{javaheri2017subjectiveB} & DSIS & 2D monitor & Passive & Point\\
      Javaheri \textit{et al.}~\cite{javaheri2017subjectiveA} & DSIS & 2D monitor & Passive & Mesh\\
      Alexiou \textit{et al.}~\cite{alexiou2017subjective} & DSIS & 2D monitor & Interactive & Point\\
      Alexiou \textit{et al.}~\cite{alexiou2017performance} & DSIS, ACR & 2D monitor & Interactive & Point\\
      Alexiou \textit{et al.}~\cite{alexiou2017towards} & DSIS & HMD (AR) & Interactive & Point\\
      Alexiou \textit{et al.}~\cite{alexiou2018pointAS} & DSIS, ACR & 2D monitor & Interactive & Point\\
      Alexiou \textit{et al.}~\cite{alexiou2018impact} & DSIS & 2D monitor, HMD (AR) & Interactive & Point\\
      Alexiou \textit{et al.}~\cite{alexiou2018benchmarking} & DSIS, ACR & 2D monitor & Interactive & Point\\
      Alexiou \textit{et al.}~\cite{alexiou2018point} & DSIS & 2D monitor & Passive & Point, Mesh\\
      Torlig \textit{et al.}~\cite{torlig2018novel} & DSIS & 2D monitor & Passive & Point\\
      Alexiou \textit{et al.}~\cite{alexious2018point} & DSIS & 3D monitor & Passive, Interactive & Mesh\\
      Zhang \textit{et al.}~\cite{zhang2014subjective} & -- & 2D monitor & -- & Point\\
      Nehmé \textit{et al.}~\cite{nehme2019comparison} & DSIS, ACR & HMD (VR) & Passive & Mesh\\
      Zerman \textit{et al.}~\cite{zerman2019subjective} & DSIS, PC & 2D monitor & Interactive & Point\\
      Alexiou \textit{et al.}~\cite{alexiou2019exploiting} & DSIS & 2D monitor & Interactive & Point\\
      da Silva Cruz \textit{et al.}~\cite{da2019point} & DSIS & 2D monitor & Passive & Point\\
      Alexiou \textit{et al.}~\cite{alexiou2019comprehensive} & DSIS & 2D monitor & Interactive & Point\\
      Javaheri \textit{et al.}~\cite{javaheri2020point} & DSIS & 2D monitor & Passive & Point, Mesh\\
      Jesús Gutiérrez \textit{et al.}~\cite{gutierrez2020quality} & ACR-(HR) & HMD (MR) & Interactive & Mesh\\
      Viola \textit{et al.}~\cite{viola2020color} & DSIS & 2D monitor & - & Point\\
      Alexiou \textit{et al.}~\cite{alexiou2020pointxr} & DSIS, ACR & 2D monitor, HMD (VR) & Passive, Interactive & Point\\
      Zerman \textit{et al.}~\cite{zerman2020textured} & ACR & 2D monitor & Passive & Point\\
      Javaheri \textit{et al.}~\cite{javaheri2020improving} & DSIS & 2D monitor & Passive & Point\\
      Hua \textit{et al.}~\cite{hua2020vqa} & - & 2D monitor & - & Point\\
      Yang \textit{et al.}~\cite{yang2020predicting} & ACR-(HR) & 2D monitor & Interactive & Point\\
      Perry \textit{et al.}~\cite{perry2020quality} & DSIS & 2D monitor & Passive & Point\\
      Liu \textit{et al.}~\cite{liu2020point} & DSIS & 2D monitor & Interactive & Point\\
      Nehmé \textit{et al.}~\cite{nehme2020visual} & DSIS & HMD (VR) & Interactive & Mesh\\
      Wu \textit{et al.}~\cite{wu2021subjective} & DSIS & HMD & Interactive & Point\\
      Ours & DSIS & 2D monitor & Passive & Point\\
  \bottomrule
  \end{tabular}
}
\end{table*}

We believe that a large-scale subject-rated PCQA database of diverse, high quality source point clouds will benefit the PCQA research community.
In this work, we first introduce our procedure of constructing a high quality color point cloud, which comprises dense points, presents flawless object surfaces, and can be viewed from any direction. 
Following the same procedure, we build so far the largest source point cloud repository, consisting of $20$ high quality point clouds with diverse geometric characteristics and textural patterns. 
By degrading each reference point cloud of the repository with $5$ types of distortion, we derive $740$ distorted point clouds, 
whose perceptual qualities are then evaluated as mean opinion scores (MOS) through a carefully designed subjective user study.
As a result, we construct currently the largest subject-rated PCQA database, termed as the Waterloo Point Cloud (WPC) database.
With the WPC database, we conduct a comprehensive evaluation on existing objective PCQA models for the first time, and find that none of these models can achieve a correlation close to an average person's performance. 
We then develop a novel objective PCQA model, inspired by the information content weighted SSIM measure (IW-SSIM)~\cite{wang2010information}. 
Experimental results show that the proposed model well correlates with subjective scores and significantly outperforms existing objective PCQA models. 

Partial and preliminary results of this work were presented in~\cite{su2019perceptual}, and the new contributions of this paper are three folds. Firstly, we provide in-depth details of the WPC database and perform new data analysis on the subjective test, which shows the diversity of the source content and the reliability of the subjective scores. Secondly, we perform a thorough comparison of existing PCQA models, revealing important implications for future directions of the development of PCQA models. Finally, we propose a new objective PCQA model, which achieves state-of-the-art performance on the proposed WPC database and other popular PCQA databases.

\section{Related Work}\label{sec:literature}
\subsection{Existing PCQA Databases}
Although a large-scale point cloud database with reliable subjective quality scores is as essential to PCQA study as the LIVE database~\cite{sheikh2005live} to the IQA research\cite{sheikh2006statistical}, acquiring high quality point clouds is not as easy as capturing high quality images with a modern camera.
A summary of publicly-available point cloud databases for different purposes can be found in the Point Cloud Library (PCL) website~\cite{rusu20113d}. 
Here we are focused on point cloud databases for quality assessment, \textit{i.e.} PCQA databases.
Since 1994, Turk and Levoy~\cite{turk1994zippered} started to build the Stanford 3D scanning repository, in which some representative point clouds, such as ``Stanford Bunny'', are still used in recent researches~\cite{alexiou2017performance,alexiou2017towards,alexiou2018point}. 
Later, MPEG point cloud database~\cite{MPEG2017Datasets} and JPEG Pleno database~\cite{JPEGPleno} collect more point clouds from multiple resources, whose contents include cultural heritages, computer-generated objects, human figures, etc. 
The three point cloud databases provide a good number of visual stimuli for a series of subjective studies on point cloud quality~\cite{javaheri2020point,zerman2019subjective,alexiou2017performance,alexiou2017towards,alexiou2018point,alexiou2019comprehensive,zerman2020textured,alexiou2020pointxr,hua2020vqa,hua2021cpc,yang2020predicting,perry2020quality,nehme2020visual,wu2021subjective}.
In fact, most point clouds used in existing publicly-available PCQA databases~\cite{javaheri2020point,zerman2019subjective,alexiou2017performance,alexiou2017towards,alexiou2018point,alexiou2019comprehensive,zerman2020textured,alexiou2020pointxr,hua2020vqa,hua2021cpc,yang2020predicting,perry2020quality,nehme2020visual,wu2021subjective} are selected from the three point cloud databases.
This makes existing PCQA databases suffer from similar shortcomings. 
First, most point clouds in the Stanford 3D scanning repository~\cite{turk1994zippered} are colorless, and thus cannot simulate distortions associated with color/attribute degradation.
Second, the scanning process does not capture the aesthetic quality of the objects, especially for cultural heritages. 
Typical examples include the ``RomanOilLight''~\cite{JPEGPleno} as shown in Fig.~\ref{fig:otherpc} (a) and the ``Head''~\cite{MPEG2017Datasets} as shown in Fig.~\ref{fig:otherpc} (b). 
Third, the perceptual quality of some point clouds are inferior due to scanning noise (``Statue\_Klimt''~\cite{MPEG2017Datasets} as shown in Fig.~\ref{fig:otherpc} (c)), and irregular edges (``Phil''~\cite{loop2016microsoft} as shown in Fig.~\ref{fig:otherpc} (d)), etc; 
Fourth, many point clouds in the three databases are scanned only from the front side, and thus cannot be watched from many viewpoints.
Fig.~\ref{fig:otherpc} (e) shows an example of such point clouds observed from invalid viewpoints.
In contrast, the human figure point cloud~\cite{d20178i} as shown in Fig.~\ref{fig:otherpc} (f) is a good source content for its excellent quality and integrity. 
Nevertheless, high quality point clouds like this are rarely found in existing PCQA databases~\cite{javaheri2020point,zerman2019subjective,alexiou2017performance,alexiou2017towards,alexiou2018point,alexiou2019comprehensive,wu2021subjective}. Fifth, existing PCQA databases exhibit relatively low content diversity.
We wish to address these limitations in the proposed WPC database by generating new high quality point cloud contents and introducing various distortion types.

The 3D nature and irregular data representation of point clouds also add new complications to the design of subjective test.
We summarize all existing subjective tests~\cite{javaheri2017subjectiveA,alexiou2017subjective,alexiou2018point,alexious2018point,javaheri2017subjectiveB,torlig2018novel,zerman2019subjective,alexiou2017performance,alexiou2017towards,alexiou2018pointAS,alexiou2018impact,alexiou2018benchmarking,zhang2014subjective,nehme2019comparison,alexiou2019exploiting,da2019point,alexiou2019comprehensive,javaheri2020point,gutierrez2020quality,viola2020color,alexiou2020pointxr,zerman2020textured,javaheri2020improving,hua2020vqa,yang2020predicting,perry2020quality,liu2020point,nehme2020visual,wu2021subjective} on PCQA in Table~\ref{tab:subjectivetest}, including both publicly-available and privately-held ones.
From the table, we find that existing subjective tests vary on the scoring methodology, the viewing display, the interaction method, and the rendering mode.
Below are several important observations.
First, compared with Absolute Category Rating (ACR) and Pairwise Comparison (PC), Double Stimulus Impairment Scale (DSIS)~\cite{rec2012bt} is more often used.
One possible reason is that most point clouds used in the subjective tests are cultural heritages or computer-generated objects, which are not common in daily life.
Therefore, subjects may not be able to detect all the distortions, especially the color impairments~\cite{nehme2019comparison}, in the ACR method without a reference.
Second, most subjective tests prefer 2D monitors to the more advanced 3D monitors and HMDs.
We speculate that the latter devices cause more 3D dizziness and sickness that may impair participants' perception of quality~\cite{solimini2013there,sharples2008virtual}.
Third, there are mainly two methods for subjects to interact with and view point clouds from 360 degrees.
The first method lets the subjects interact freely with the displayed point cloud using a mouse/head-mounted display (HMD), while the other one rotates the point cloud along a predetermined virtual path so that subjects can observe the point cloud from various viewpoints and distances.
Since existing subjective tests show little bias on the interaction method, implying that both methods work similarly well, the passive method has an edge in repeatability and reproducibility over the interactive one.
Fourth, the point-based rendering method is overwhelmingly popular in existing subjective tests due to its simplicity and low computational complexity.
While there already exist various subjective test settings in the literature, which setting is the best remains an open-ended question.
In this paper, we will adopt the most common settings, as shown in the last line of Table~\ref{tab:subjectivetest}, to conduct the subjective test.

\subsection{Objective Quality Assessment of 3D Point Clouds}
A handful of algorithms have been proposed for PCQA, which can be roughly categorized as follows. 
From the perspective of distortion type, objective PCQA models can be classed as geometry distortion metrics~\cite{tian2017geometric,mekuria2017design,mekuria2017performance,tian2017evaluation,alexiou2018pointAS,cignoni1998metro,meynet2019pc,javaheri2020generalized,javaheri2020improving,javaheri2020mahalanobis} and geometry-plus-color distortion metrics~\cite{mekuria2017design,mekuria2017performance,alexiou2020towards,meynet2020pcqm,viola2020color,diniz2020towards,diniz2020multi,hua2020vqa,hua2021cpc,yang2020inferring,torlig2018novel,alexiou2019exploiting,alexiou2019comprehensive,su2019perceptual,yang2020predicting,diniz2021novel,diniz2020local,wu2021subjective,he2021towards,diniz2021color,liu2020point,hua2021bqe,tao2021point,xu2021epes,zhang2021ms}.
From the perspective of feature extraction, objective PCQA models can be classed as point-based models~\cite{tian2017geometric,mekuria2017design,mekuria2017performance,alexiou2018pointAS,meynet2019pc,javaheri2020generalized,javaheri2020improving,javaheri2020mahalanobis,alexiou2020towards,meynet2020pcqm,viola2020color,diniz2020towards,diniz2020multi,hua2020vqa,hua2021cpc,yang2020inferring,diniz2021novel,diniz2020local,diniz2021color,liu2020point,hua2021bqe,xu2021epes,zhang2021ms} and projection-based models~\cite{torlig2018novel,alexiou2019exploiting,alexiou2019comprehensive,su2019perceptual,yang2020predicting,wu2021subjective,he2021towards,tao2021point}. 

More specifically, point-to-point and point-to-plane metrics, both adopted by the MPEG PCC group, assess geometric distortions by Euclidean distances or projected errors along normal vector directions~\cite{tian2017geometric,mekuria2017design,mekuria2017performance}. In~\cite{alexiou2018pointAS}, angular similarity metrics estimate the geometric distortions by calculating the similarity of local surface approximations. To obtain the geometric distortion, the PC-MSDM model~\cite{meynet2019pc} computes curvature similarities between original and distorted PCs. However, because PC normals are very sensitive to different normal acquisition methods, normals in the above metrics~\cite{tian2017geometric,mekuria2017design,mekuria2017performance,alexiou2018pointAS,meynet2019pc} may lead to uncertain performance. The generalized Hausdorff
distance metric~\cite{javaheri2020generalized} exploits multiple rankings to identify the best performing quality model in terms of correlation
with subjective quality, overcoming the oversensitivity of the classical Hausdorff distance to outliers. In~\cite{javaheri2020improving}, a PSNR-based metric~\cite{tian2017geometric,mekuria2017design,mekuria2017performance} is proposed by including a normalization factor that accounts for changes in the intrinsic PC resolution after rendering. In~\cite{mekuria2017performance}, PSNR-based methods are modified by a density coefficient determined by the coordinate peak and the rendering resolution. In~\cite{javaheri2020mahalanobis}, a point-to-distribution quality assessment model is proposed by exploiting the correspondence between a point and a distribution of points from a small PC region. The main idea in this paper is to characterize the PC surface through the covariance
of points within some local region, which is not extremely influenced by the number of reconstructed points after decoding, but rather by a statistical characterization of the point locations.

Compared with geometry-only PCs, colored PCs have a broad range of applications. Many point-based metrics for colored PCs have emerged recently~\cite{mekuria2017design,mekuria2017performance,alexiou2020towards,meynet2020pcqm,viola2020color,diniz2020towards,diniz2020multi,hua2020vqa,yang2020inferring,hua2021cpc,viola2020reduced,liu2021reduced,diniz2021color,diniz2020local,diniz2021novel,hua2021bqe,xu2021epes,zhang2021ms}. In~\cite{mekuria2017design,mekuria2017performance}, point-to-point PSNR on the Y component (MPEG PSNR$_{Y}$) is used to estimate texture distortion of colored PCs, though such a direct extension of PSNR inevitably inherits the widely-known disadvantages of PSNR. Similarity measures~\cite{wang2004image} proven to be effective in general image quality assessment are extended to PCQA~\cite{alexiou2020towards,meynet2020pcqm,hua2020vqa}. In these methods, geometry-based, color-based, normal-based and curvature-based features are extracted from both reference and distorted PCs, then both geometry and color feature similarities are evaluated and combined to overall objective scores. In~\cite{viola2020color}, color histograms and correlograms are used to estimate the impairment of a distorted PC with respect to its reference. Geometry-only and color-only approaches are then combined to get a rendering-independent objective PCQA metric. More recently, statistics of a variant of the Local Binary Pattern (LBP)~\cite{diniz2020towards,diniz2020multi}, Perceptual Color Distance Patterns (PCDP)~\cite{diniz2021novel} and Local Luminance Patterns (LLP)~\cite{diniz2020local} descriptors are introduced to PCQA. In \cite{diniz2021color}, the BitDance metric uses color and geometry texture descriptors. The proposed method first extracts the statistics of color and geometry information of the reference and test PCs. Then, it compares the color and geometry statistics and combines them to estimate the perceived quality of the test PC. The GraphSIM approach\cite{yang2020inferring,zhang2021ms} uses graph signal gradient as a quality index to evaluate PC distortions. Considering the visual masking effect of PC’s geometric information and the color perception of human eyes, the CPC-GSCT metric~\cite{hua2021cpc} uses geometric segmentation and color transformation respectively to construct geometric and color features and then to estimate the PC quality. Similarly, the BQE-CVP metric~\cite{hua2021bqe} uses geometric feature, color feature and joint featrue to develop a blind quality evaluator. In~\cite{viola2020reduced}, a reduced reference PCQA metric is developed that extracts geometry-based, normal-based and luminance-based features from the reference PC. Such features are then transmitted alongside the content, and are employed at the receiver side and compared with the distorted PC. The best combination of the features is obtained through a linear optimization algorithm. In~\cite{liu2021reduced}, two color features are proposed to estimate three content dependent parameters, and then a reduced reference PCQA model is established. The reason is that the content has a masking effect on the coding distortion that is consistent with the characteristics of the human visual system. That is to say, the parameters are highly content dependent. 

In addition to the aforementioned point-based models, there are also many projection-based models~\cite{su2019perceptual,torlig2018novel,alexiou2019exploiting,alexiou2019comprehensive,yang2020predicting,wu2021subjective,he2021towards,tao2021point}, such as projection-based PSNR (PSNRp)~\cite{su2019perceptual,torlig2018novel}, projection-based structural similarity (SSIMp)~\cite{wang2004image,su2019perceptual,torlig2018novel}, projection-based multi-scale structural similarity (MS-SSIMp)~\cite{wang2003multiscale,su2019perceptual,torlig2018novel} and projection-based pixel-domain visual information fidelity (VIFPp)~\cite{Sheikh2006image,su2019perceptual,torlig2018novel}. Yang \textit{et al.}~\cite{yang2020predicting} choose to project the 3D point cloud onto six perpendicular image planes of a cube for the color texture image and corresponding depth image, and aggregate image-based global and local features among all projected planes for a final objective index. Wu \textit{et al.}~\cite{wu2021subjective} propose two projection-based objective quality
evaluation methods: a weighted view projection based model and a patch projection based model. He \textit{et al.}~\cite{he2021towards} propose a method combining colored texture and curvature projection. Specifically, the colored texture information and curvature of PC are projected onto 2D planes to extract texture and geometric statistical features, respectively, so as to characterize the texture and geometric distortion. However, these methods treat background padding pixels on projected image planes the same way as the foreground ones, leading to inferior quality prediction accuracy~\cite{wang2010information}. Alexiou \textit{et al.}~\cite{alexiou2019exploiting,alexiou2019comprehensive} develop an additional algorithm to remove the influence of background pixels, but the model complexity increases and the robustness declines. 

\section{Point Cloud Database Construction}\label{sec:database}
\begin{figure*}[htb]
    \centering
    \subfloat[]{\includegraphics[width=0.095\textwidth]{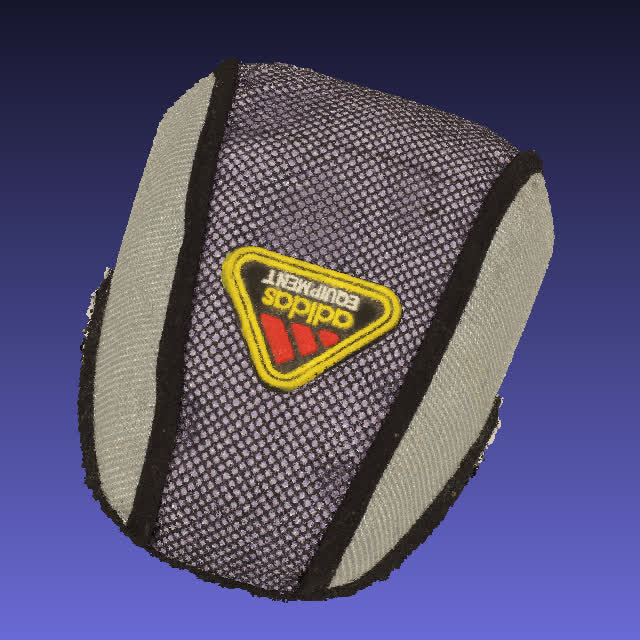}}\hskip0.1em
    \subfloat[]{\includegraphics[width=0.095\textwidth]{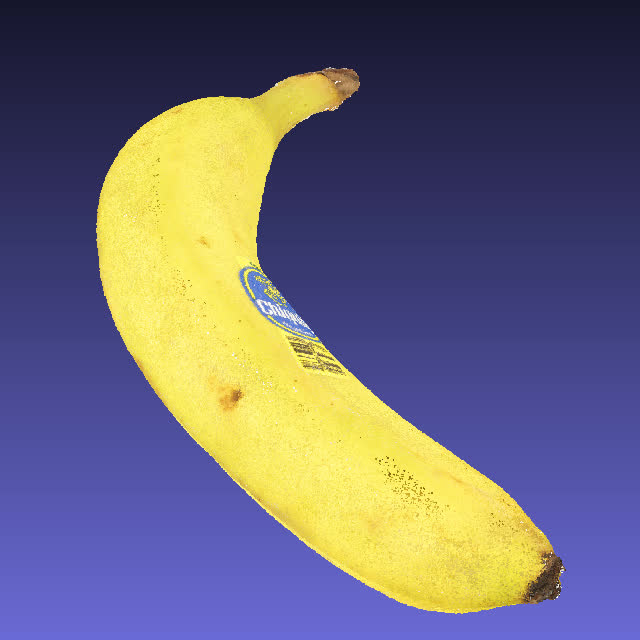}}\hskip0.1em
    \subfloat[]{\includegraphics[width=0.095\textwidth]{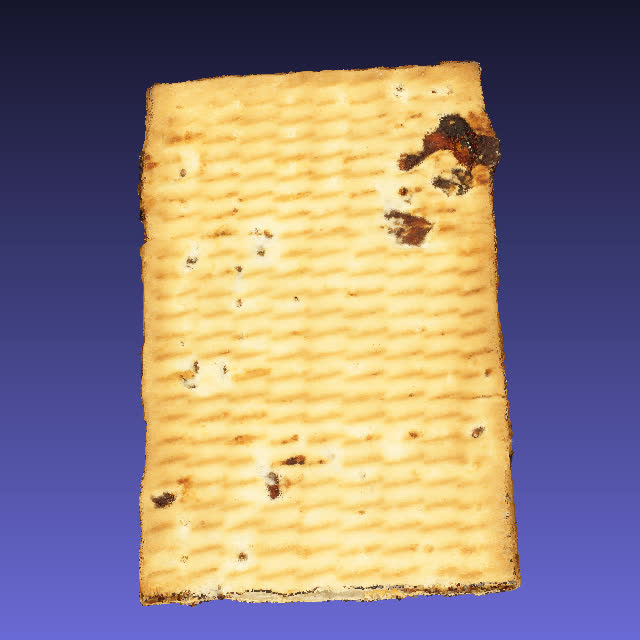}}\hskip0.1em
    \subfloat[]{\includegraphics[width=0.095\textwidth]{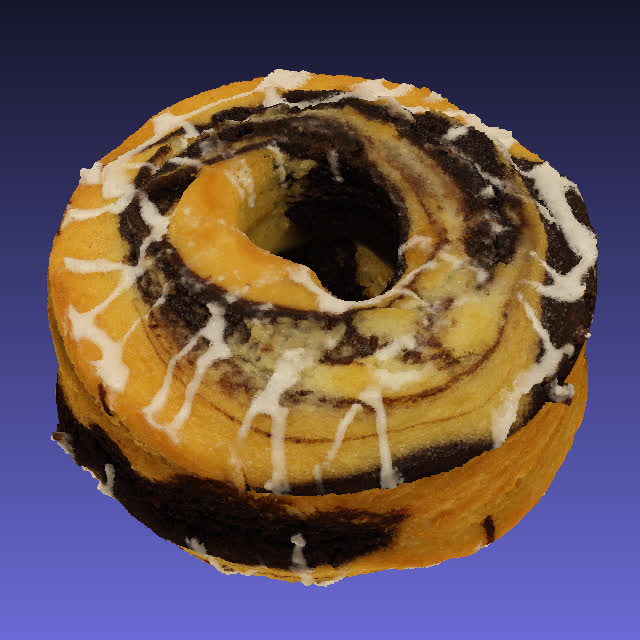}}\hskip0.1em
    \subfloat[]{\includegraphics[width=0.095\textwidth]{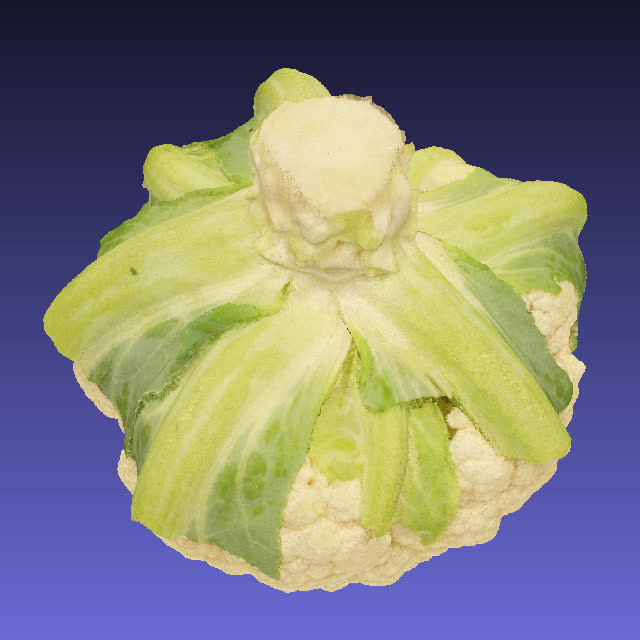}}\hskip0.1em
    \subfloat[]{\includegraphics[width=0.095\textwidth]{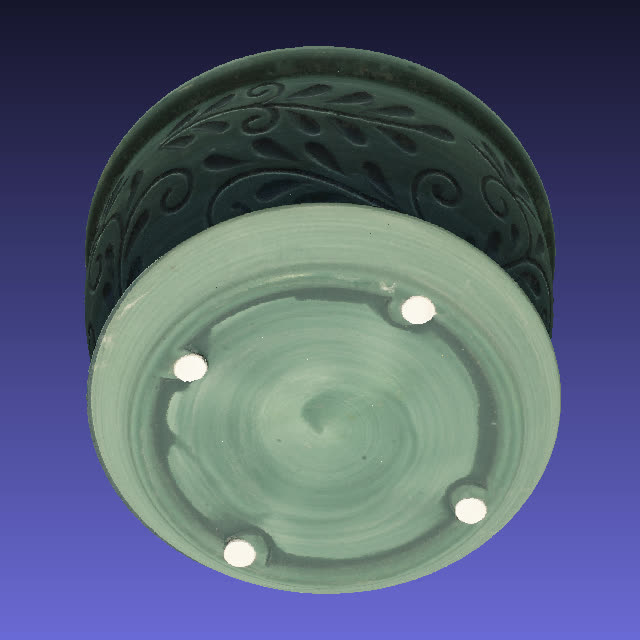}}\hskip0.1em
    \subfloat[]{\includegraphics[width=0.095\textwidth]{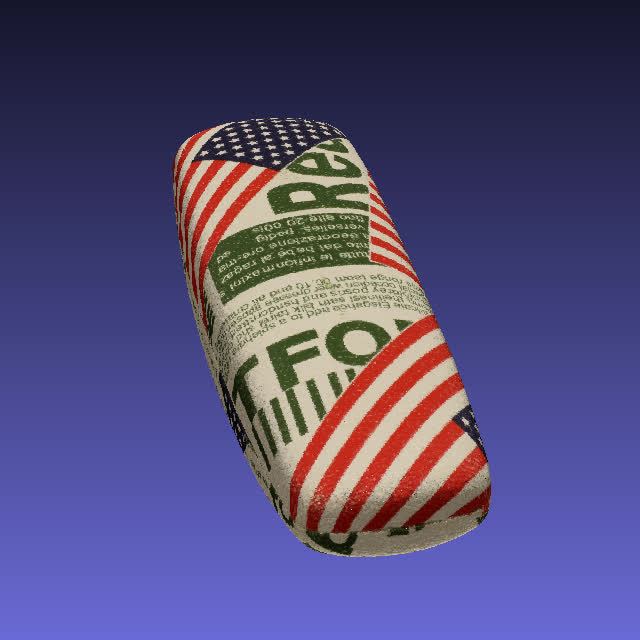}}\hskip0.1em
    \subfloat[]{\includegraphics[width=0.095\textwidth]{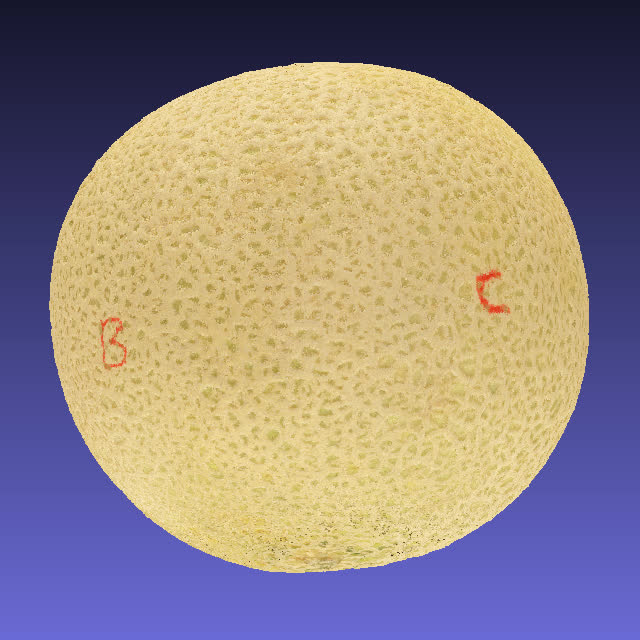}}\hskip0.1em
    \subfloat[]{\includegraphics[width=0.095\textwidth]{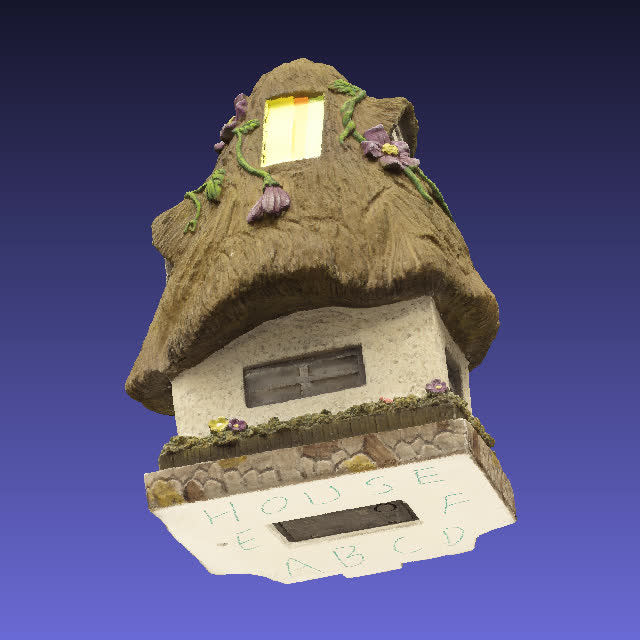}}\hskip0.1em
    \subfloat[]{\includegraphics[width=0.095\textwidth]{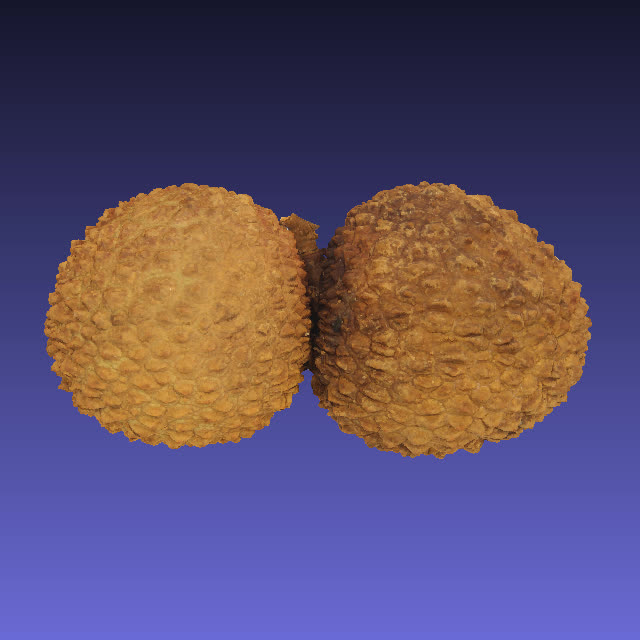}}\hskip0.1em
    \subfloat[]{\includegraphics[width=0.095\textwidth]{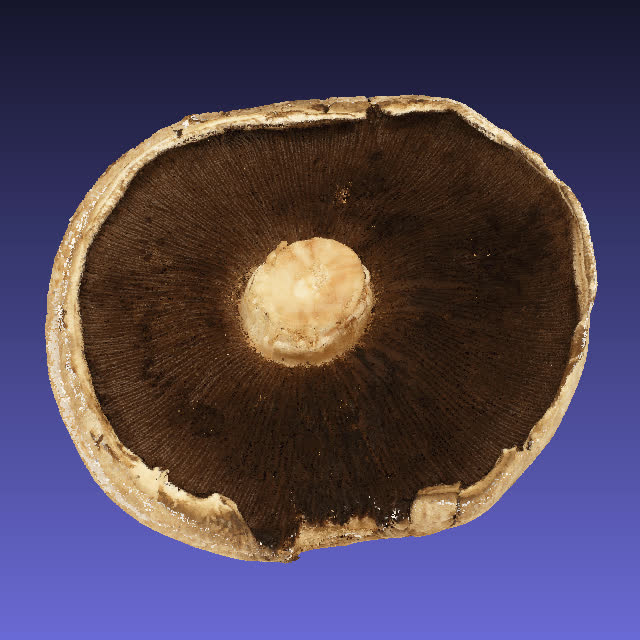}}\hskip0.1em
    \subfloat[]{\includegraphics[width=0.095\textwidth]{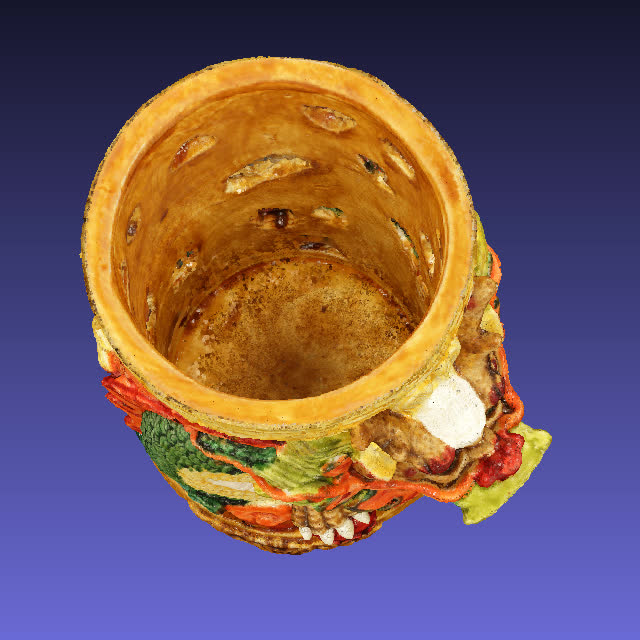}}\hskip0.1em
    \subfloat[]{\includegraphics[width=0.095\textwidth]{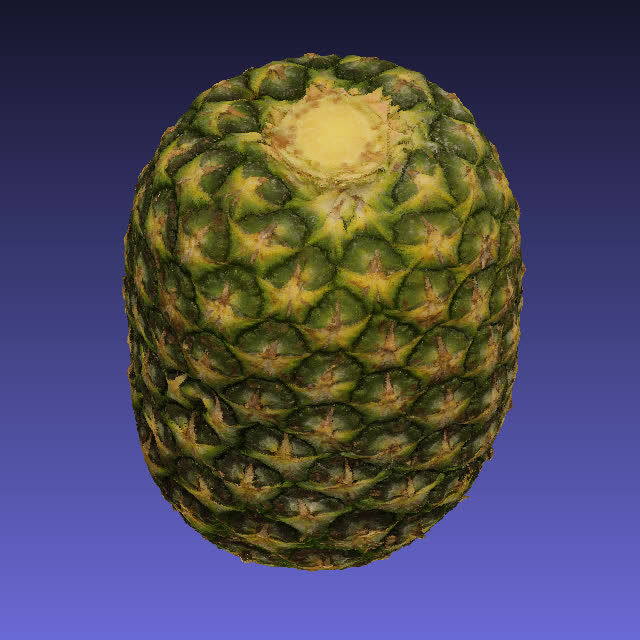}}\hskip0.1em
    \subfloat[]{\includegraphics[width=0.095\textwidth]{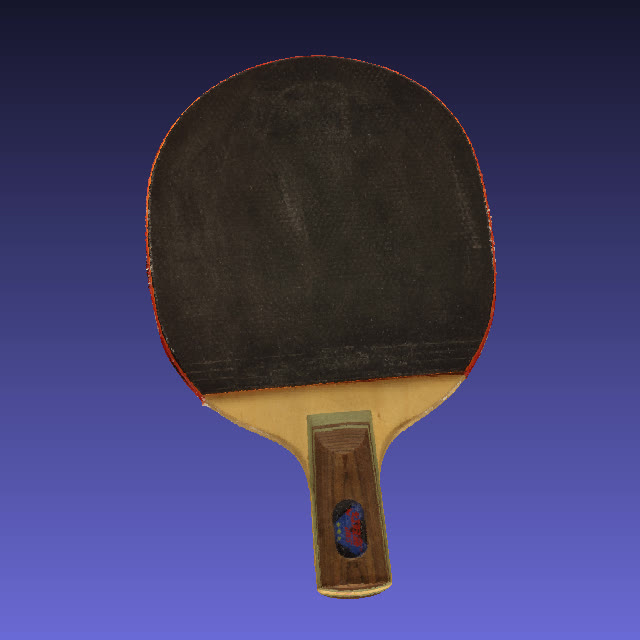}}\hskip0.1em
    \subfloat[]{\includegraphics[width=0.095\textwidth]{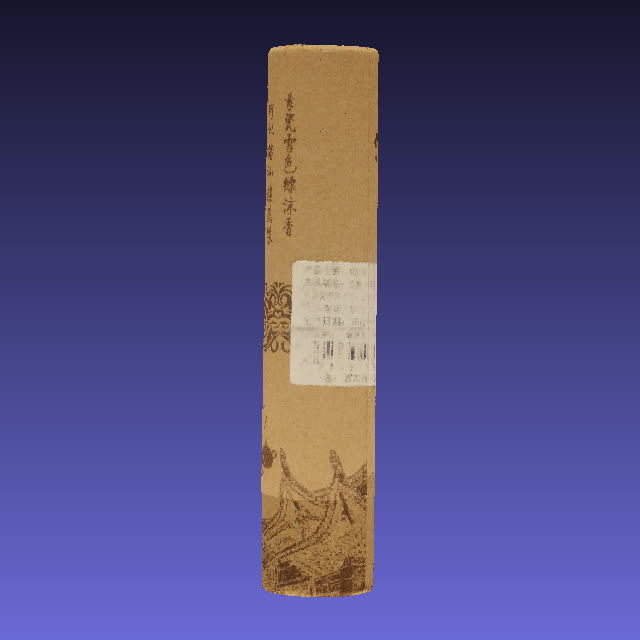}}\hskip0.1em
    \subfloat[]{\includegraphics[width=0.095\textwidth]{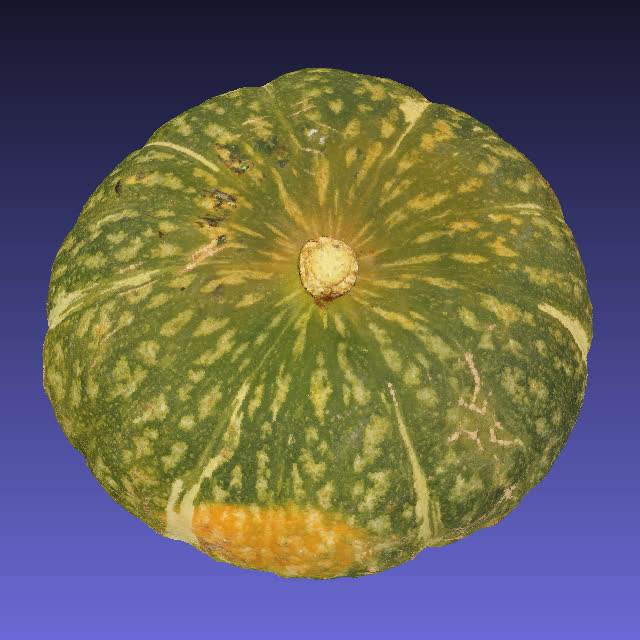}}\hskip0.1em
    \subfloat[]{\includegraphics[width=0.095\textwidth]{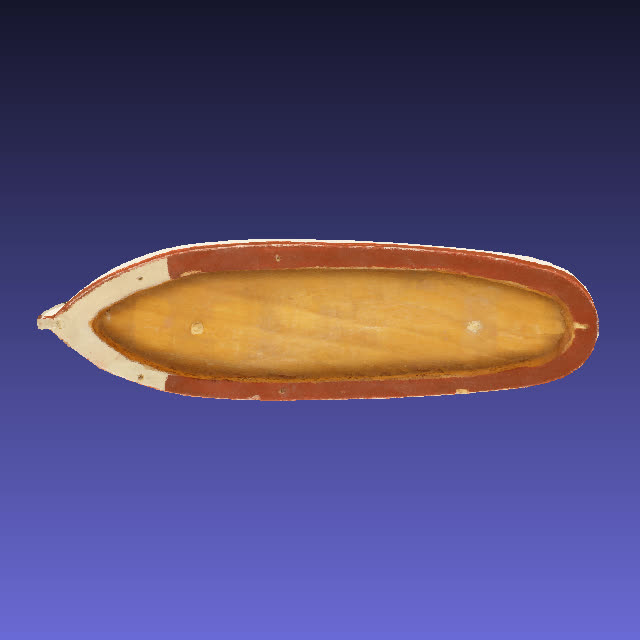}}\hskip0.1em
    \subfloat[]{\includegraphics[width=0.095\textwidth]{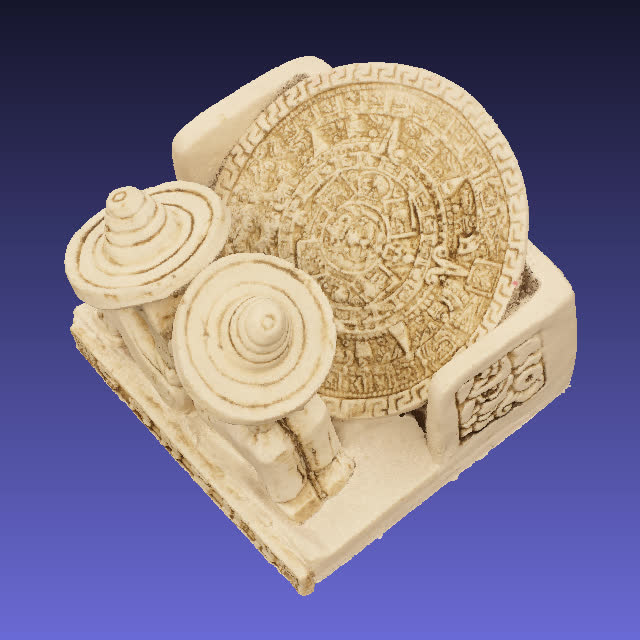}}\hskip0.1em
    \subfloat[]{\includegraphics[width=0.095\textwidth]{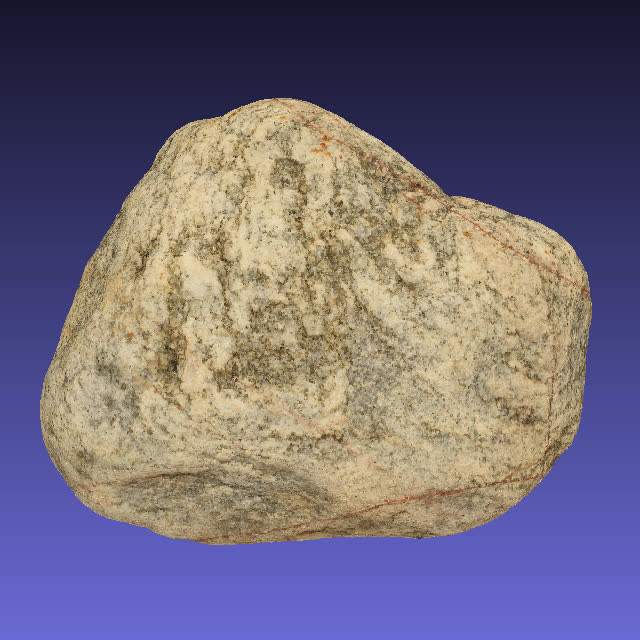}}\hskip0.1em
    \subfloat[]{\includegraphics[width=0.095\textwidth]{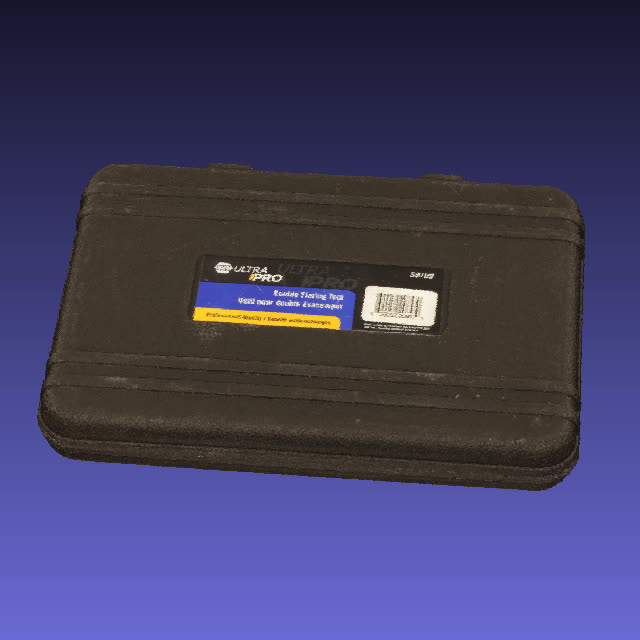}}\hskip0.1em
    \caption{Snapshots of acquired point clouds in the Waterloo Point Cloud database.}
    \label{fig:sourcepc}
\end{figure*}

\begin{figure}[t!]
    \centering
    \includegraphics[width=\linewidth]{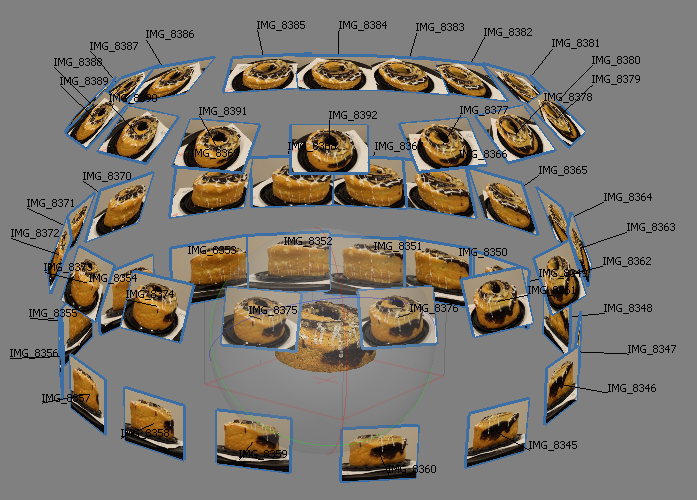}
    \caption{Sample point cloud acquisition process.}
    \label{fig:acquisition}
\end{figure}
\subsection{Point Cloud Construction}

Motivated by the lack of source 3D point cloud, we gather a collection of objects with diverse geometric and textural complexity, including snacks, fruits, vegetables, office supplies, and containers, etc. The selected contents are moderate in size and are omni-directionally acquirable. Fig.~\ref{fig:sourcepc} shows snapshots of the objects in our point cloud dataset. The 3D point clouds are constructed using the following steps.

\begin{figure*}[t]
    \centering
    \subfloat[]{\includegraphics[width=0.195\textwidth]{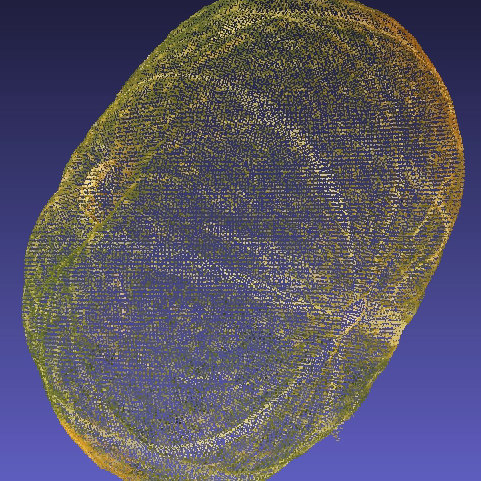}}\hskip0.1em
    \subfloat[]{\includegraphics[width=0.195\textwidth]{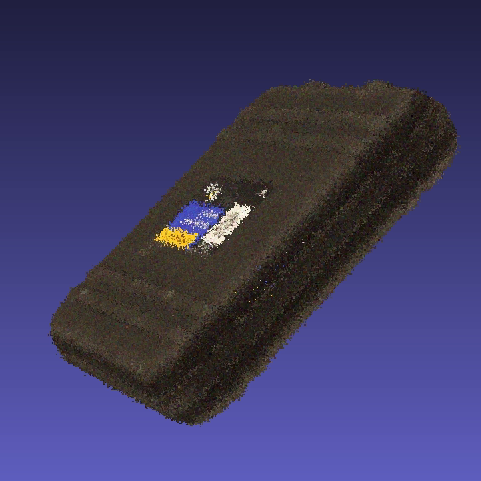}}\hskip0.1em
    \subfloat[]{\includegraphics[width=0.195\textwidth]{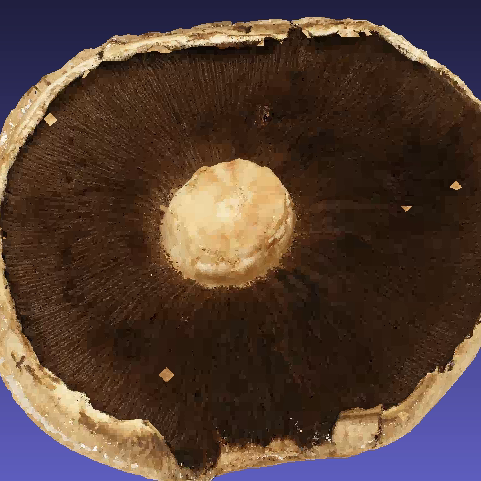}}\hskip0.1em
    \subfloat[]{\includegraphics[width=0.195\textwidth]{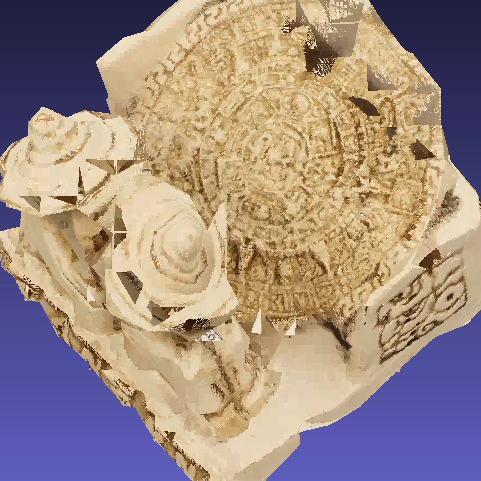}}\hskip0.1em
    \subfloat[]{\includegraphics[width=0.195\textwidth]{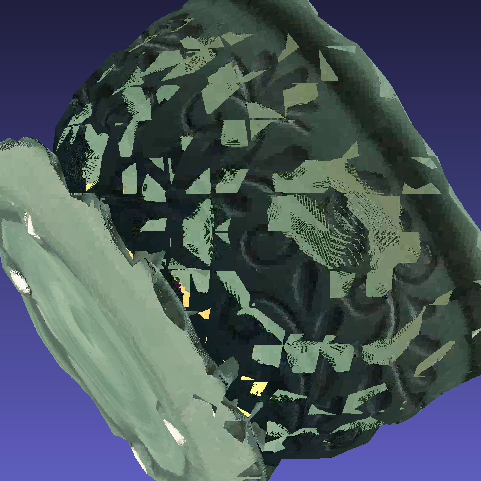}}
    
    \subfloat[]{\includegraphics[width=0.195\textwidth]{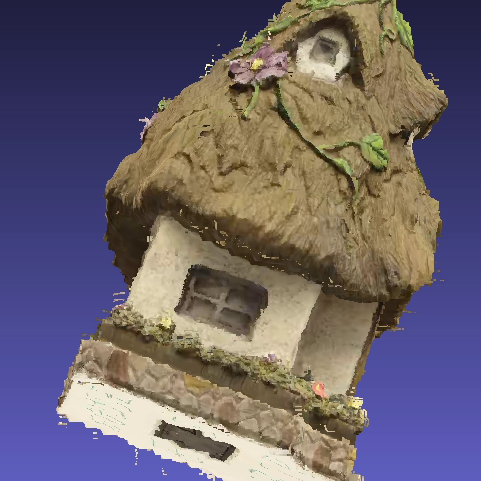}}\hskip0.1em
    \subfloat[]{\includegraphics[width=0.195\textwidth]{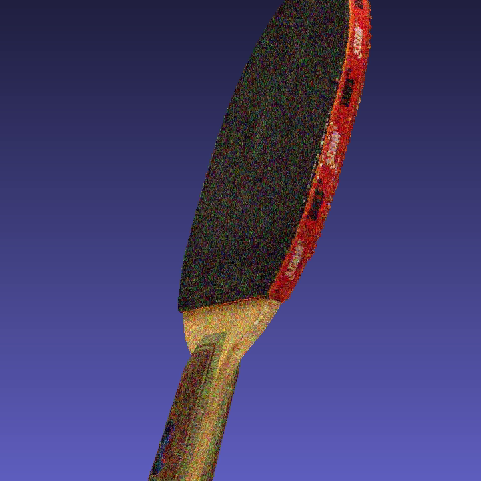}}\hskip0.1em
    \subfloat[]{\includegraphics[width=0.195\textwidth]{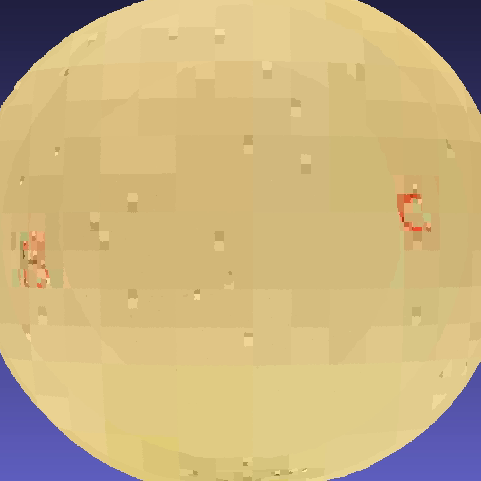}}\hskip0.1em
    \subfloat[]{\includegraphics[width=0.195\textwidth]{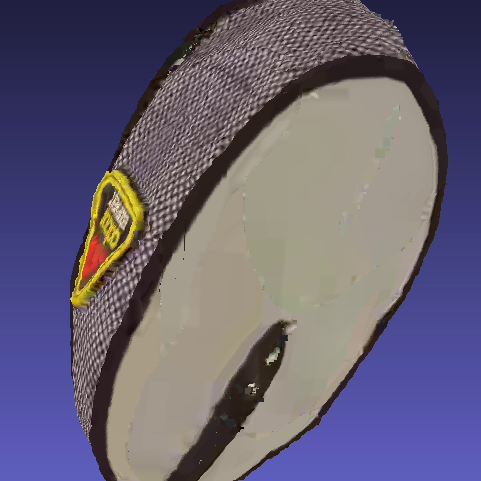}}\hskip0.1em
    \subfloat[]{\includegraphics[width=0.195\textwidth]{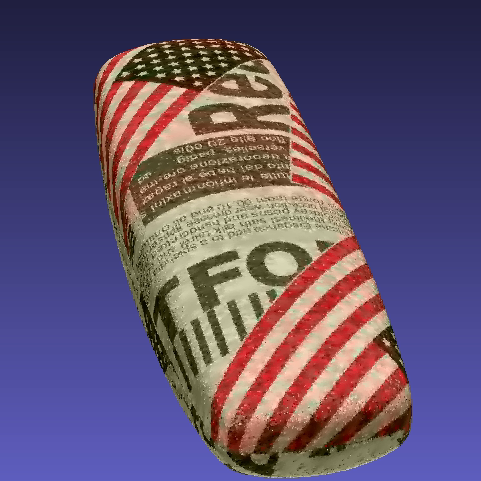}}

    \caption{Point cloud distortions. Geometry distortions: (a) Hollow. (b) Geometry noise. (c) Hole. (d) Shape distortion. (e) Collapse. (f) Gap and burr. Texture distortions: (g) Texture noise. (h) Blockiness. (i) Blur. (j) Color bleeding.}
    \label{fig:distortion}
\end{figure*}

\begin{table*}[t]
\centering
\caption{Characteristics of acquired point clouds in the Waterloo Point Cloud database. X, Y, Z: coordinates in three dimensions, DNN: distance between nearest neighbors, GC: geometric complexity, and TC: textural complexity.}
\label{tab:sourcepc}
\scalebox{0.8}{
  \begin{tabular}{c|c|c|c c c c|c c|c}
  \toprule
      Index & Name & Points & Xmin, Ymin, Zmin & Xmax & Ymax & Zmax & Min DNN & Max DNN & Description\\\hline
      a & Bag & 1267845 & 0 & 879 & 1000 & 605 & 1 & 8.94 & Daily supply, high TC\\
      b & Banana & 807184 & 0 & 828 & 1000 & 900 & 1 & 18.47 & Fruit, low TC\\
      c & Biscuits & 952579 & 0 & 693 & 1000 & 631 & 1 & 10.05 & Snack, thin, medium TC\\
      d & Cake & 2486566 & 0 & 1000 & 953 & 970 & 1 & 7.14 & Snack, topological hole, medium TC\\
      e & Cauliflower & 1936627 & 0 & 1000 & 964 & 956 & 1 & 10.20 & Vegetable, low TC\\
      f & Flowerpot & 2407154 & 0 & 896 & 1000 & 950 & 1 & 15.65 & Container, thin wall, low TC\\
      g & Glasses\_case & 716659 & 0 & 579 & 798 & 1000 & 1 & 24.37 & Daily supply, high TC\\
      h & Honeydew\_melon & 1431071 & 0 & 984 & 928 & 1000 & 1 & 29.97 & Fruit, medium TC\\
      i & House & 1568490 & 0 & 717 & 1000 & 755 & 1 & 8.77 & Crafts, high GC, high TC\\
      j & Litchi & 1039942 & 0 & 1000 & 510 & 550 & 1 & 2.45 & Fruit, medium TC\\
      k & Mushroom & 1144603 & 0 & 1000 & 857 & 568 & 1 & 18.60 & Vegetable, thin, different GC and TC on both sides\\
      l & Pen\_container & 2878318 & 0 & 829 & 912 & 1000 & 1 & 23.58 & Office supply, thin, high GC and different TC on both sides\\
      m & Pineapple & 1628910 & 0 & 733 & 949 & 1000 & 1 & 17.95 & Fruit, high TC\\
      n & Ping-pong\_bat & 703879 & 0 & 649 & 1000 & 400 & 1 & 5.10 & Sports equipment, thin, different GC and TC on both sides\\
      o & Puer\_tea & 412009 & 0 & 213 & 1000 & 230 & 1 & 6.71 & Container, medium TC\\
      p & Pumpkin & 1340343 & 0 & 1000 & 934 & 756 & 1 & 3.74 & Vegetable, high TC\\
      q & Ship & 684617 & 0 & 1000 & 288 & 375 & 1 & 3.61 & Crafts, high GC, low TC\\
      r & Statue & 1637577 & 0 & 948 & 1000 & 819 & 1 & 52.20 & Crafts, high GC, different TC on both sides\\
      s & Stone & 1086453 & 0 & 1000 & 815 & 586 & 1 & 75.77 & Collection, high TC\\
      t & Tool\_box & 1054211 & 0 & 1000 & 599 & 576 & 1 & 3.32 & Container, low TC\\
  \bottomrule
  \end{tabular}
}
\end{table*}

\begin{itemize}
    \item \textit{Image acquisition}: Image acquisition is conducted in standardized laboratory environment which has a normal lighting condition without reflecting ceiling walls and floor. A single-lens-reflex camera and a turntable are employed to take photos of an object from a variety of perspectives. A graph illustration of acquisition process is shown in Fig.~\ref{fig:acquisition}, where each photo is placed at its capture position relative to the object in the center. 
    \item \textit{3D reconstruction}: We apply image alignment, sparse point cloud reconstruction, dense point cloud reconstruction and point cloud merging to each sequence of images with Agisoft Photoscan~\cite{agisoft2014agisoft}. The resulting point clouds are further refined by Screened Poisson Surface Reconstruction~\cite{kazhdan2013screened} and resampling using CloudCompare~\cite{girardeau2011cloudcompare}.
    \item \textit{Normalization}: Each point cloud is normalized to a unit-cube with a step size of 0.001, where duplicated points are removed~\cite{girardeau2011cloudcompare}. Finally, 20 voxelized point clouds are generated where the number of points range between 400K to 3M, with an average of 1.35M points and a standard deviation of 656K, respectively. Detailed specifications are given in Table~\ref{tab:sourcepc}.
\end{itemize}

\subsection{Distortion Generation}
To test the capability of PCQA models in real-world applications, we chose to distort the source point clouds with the following processes.
\begin{itemize}
    \item \textit{Downsampling}: We apply octree-based downsampling~\cite{girardeau2011cloudcompare} to the normalized point clouds. Each dimension is uniformly divided into $2^{N}$ intervals, where $N$ represents the octree level. Then points located in the same cube are merged into one node. In this study, we set $N$ to be 7, 8, and 9, respectively, to cover diverse spatial resolutions.
    \item \textit{Gaussian noise contamination}: White Gaussian noise is added independently to both geometry and texture elements with standard deviation of $\left\{0, 2, 4\right\}$ and $\left\{8, 16, 32\right\}$, respectively. Then both geometry and texture elements are rounded to the nearest integer, followed by points removal by Meshlab~\cite{cignoni2008meshlab}.
    \item \textit{MPEG-PCC}: In 2017, MPEG issued a call for proposals on PCC methods for International Organization for Standardization~\cite{schwarz2018emerging}. Two technologies were chosen as test models: G-PCC for static content and dynamically capturing, and V-PCC for dynamic content, respectively. In this work, G-PCC (Trisoup) reference codec~\cite{tmc13} is employed to encode the original point clouds with `max\_NodeSizeLog$_{2}$' of $\left\{10\right\}$, `NodeSizeLog$_{2}$' of $\left\{2, 4, 6\right\}$ and `rahtQuantizationStep' of $\left\{64, 128, 256, 512\right\}$, respectively. G-PCC (Octree)~\cite{tmc3} employs downsampling method to encode the geometry information, and is thus not performed redundantly. We set the `quantizationSteps' of texture encoding as $\left\{16, 32, 48, 64\right\}$. V-PCC reference codec~\cite{tmc2} is employed to encode the original point clouds at three `geometryQP' values and three `textureQP' values, ranging from 35-50 and 35-50, respectively, followed by duplicated points removal~\cite{cignoni2008meshlab}. 
\end{itemize}
Eventually, 740 distorted point clouds are generated in total by 5 distortion generators from 20 original point clouds. In total, there are 760 original and distorted point clouds in the WPC database.

Sampled distortion patterns in the WPC database are shown in Fig.~\ref{fig:distortion}. It is interesting to observe that the distorted point cloud not only exhibits loss of texture information similar to 2D images such as blockiness and blur, but also novel geometric distortion types. Specifically, hollow is caused by point cloud downsampling, where the point density is not sufficient to cover the object surface. Holes and collapses arise from unsuccessful triangulations and inappropriate downsampling in G-PCC (Trisoup), respectively. Even when the triangulation is successful, geometry distortion may still appear as a consequence of ill-conditioned triangles. A sample case is given in the bottom right part of Fig.~\ref{fig:distortion} (d). Moreover, a large `geometryQP' in V-PCC potentially results in gaps and burrs. All these distortions are point cloud-specific, which create new challenges to objective PCQA models.

\section{Subjective experiments}\label{sec:subjective}

\subsection{Subjective User Study}

\begin{figure}[t]
  \centering
  {\includegraphics[width=1\columnwidth]{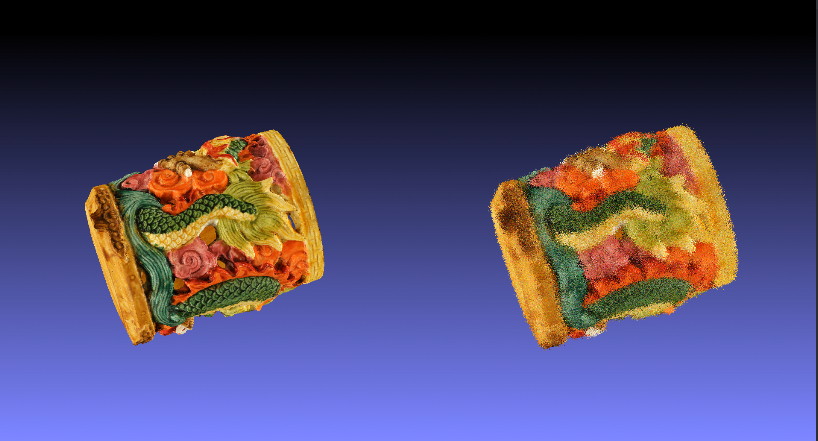}}
  \caption{A screenshot presenting ``PenContainer'' and its distorted version.}\label{fig:DSIS}
\end{figure}

We choose passive watching instead of interactive watching for subjective tests because the latter creates large variations and inconsistencies in terms of the viewpoints and viewing time between subjects and viewing sessions. We employ Technicolor renderer~\cite{renderer} to render each point cloud to a video sequence. The rendering window, point size and point type parameters are set to 960$\times$960, 1 and `point', respectively. A horizontal and a vertical circle both with a radius of 5,000 are selected successively as the virtual camera path with the center of circles at the geometry center of an object. The remaining parameters are set as default. These settings preserves detail information as much as possible while maintaining the original point clouds to be watertight. One viewpoint is generated every two degrees on these circles, resulting in 360 image frames for each point cloud. Each distorted clip is then concatenated horizontally with its pristine counterpart into a 10-second video sequence for presentation. A screenshot is shown in Fig.~\ref{fig:DSIS}.

\begin{figure}[t]
  \centering
  \subfloat[]{\includegraphics[width=0.85\columnwidth]{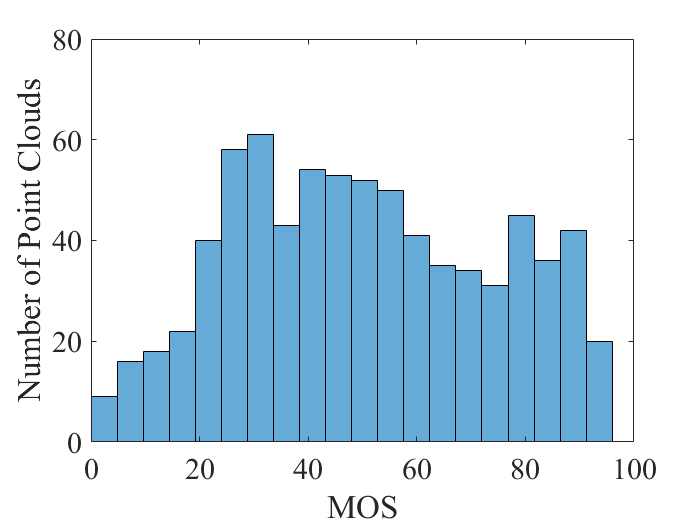}}
  
  \subfloat[]{\includegraphics[width=0.85\columnwidth]{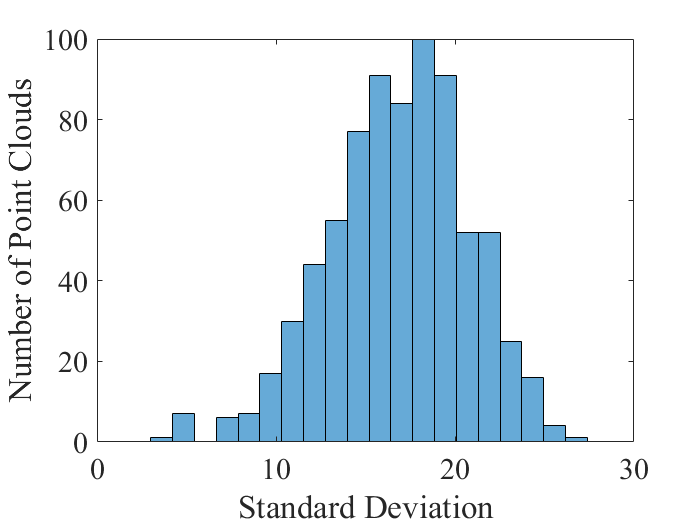}}
  \caption{MOS statistics of Waterloo Point Cloud database.}\label{fig:MOS}
\end{figure}
Our subjective testing environment is the same as that for image acquisition. All video sequences are displayed on a 23.6" LCD monitor at a resolution of 1920$\times$1080 with Truecolor (32bit) at 60 Hz. The monitor is calibrated in accordance with ITU-R Recommendation BT.500-13~\cite{rec2012bt}. Double-stimulus impairment scale (DSIS) methodology is applied in our subjective test~\cite{rec2012bt}. Videos are displayed in random order using a customized graphical user interface, where subjective scores of individual viewers are recorded. 

A total of 60 na\"ive subjects, including 32 males and 28 females aged between 21 and 40, participated in the subjective test. All the subjects have normal or corrected-to-normal vision, and viewed videos from a distance of twice the screen height. Before the testing session, a training session is performed during which 18 videos that are different from the videos in the testing session are shown to the subjects. The same methods are applied to generate videos used in both the training and testing sessions. Therefore, subjects knew what distortion types and levels to expect before the testing session, and thus the learning effects are kept minimal. Considering the limited subjective experiment capacity, we employed the following strategy. Each subject is assigned 10 objects in a circular fashion. Specifically, if subject~\textit{i} is assigned objects 1 to 10, then subject~\textit{i} + 1 watch objects 2 to 11. Each video is scored for 30 times, and 22,800 subjective ratings, including 600 scores for reference point clouds, are collected in total. For each subject, the whole study takes about 2 hours, which is divided into 4 sections with three 5-minute breaks in-between to minimize the influence of fatigue effect. For finer distinctions between ratings, 100-point continuous scale is utilized instead of a discrete 5-point ITU-R Absolute Category Scale (ACR). 
\subsection{Subjective Data Analysis}

\begin{figure}[t]
  \centering
  \subfloat[]{\includegraphics[width=0.95\columnwidth]{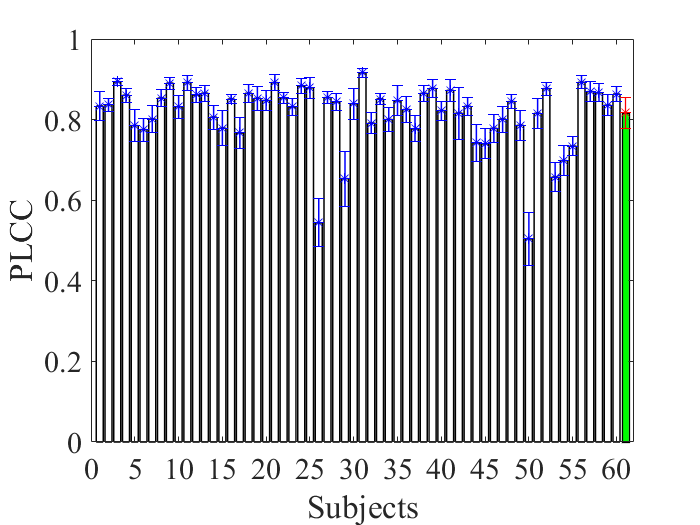}}
  
  \subfloat[]{\includegraphics[width=0.95\columnwidth]{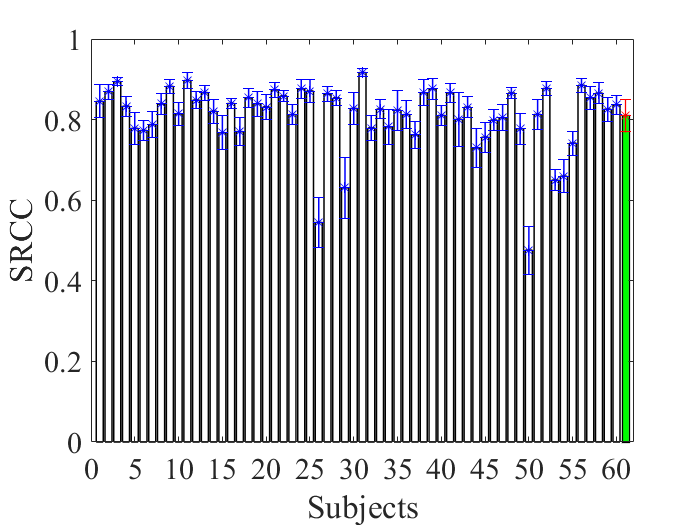}}
  \caption{PLCC and SRCC between individual subject rating and MOS. Rightmost column: performance of an average subject.}\label{fig:plccsrcc}
\end{figure}

After converting subjective scores to Z-scores, we apply the outlier removal scheme suggested in~\cite{rec2012bt}. No outlier detection is conducted participant-wise. Then Z-scores are linearly rescaled to lie in the range of [0, 100]. Mean opinion score (MOS) for each distorted point cloud is calculated by averaging the re-scaled Z-scores from all valid subjects. The histograms for the MOS and the associated standard deviation are shown in Fig.~\ref{fig:MOS}, which demonstrates that the distorted point clouds span most of the quality range. Considering the MOS as the ``ground truth'', the performance of individual subjects can be evaluated by calculating the correlation coefficient between individual subject ratings and MOS values for each source point cloud, and then averaging the correlation coefficients of all source point clouds. Pearson linear correlation coefficient (PLCC) and Spearman rank-order correlation coefficient (SRCC) are employed as the evaluation criteria. Both criteria range from 0 to 1, where higher values denote better performance. Fig.~\ref{fig:plccsrcc} depicts the mean and standard deviation of the results. We can see that most individual subjects perform quite consistently with relatively low variations for different source point cloud. The average performance across all individual subjects is also given in the rightmost columns of Fig.~\ref{fig:plccsrcc}.
\begin{table*}[t]
\centering
\caption{PLCC performance evaluation of the proposed model against existing models.}
\label{tab:plcc}
\scalebox{0.75}{
  \begin{tabular}{c|c c c c| c c c c c c c c c|c}

  \toprule
  \multirow{2}{*}{Subset}&
    \multicolumn{4}{c|}{Geometry distortion metric}&\multicolumn{9}{c|}{Geometry-plus-color distortion metric}& \multirow{2}{*}{IW-SSIM$_\emph{p}$}\\
    \cline{2-5} \cline{6-14} 
   
   & PSNR$_\emph{p2po,M}$ & PSNR$_\emph{p2po,H}$ & PSNR$_\emph{p2pl,M}$ & PSNR$_\emph{p2pl,H}$ & PSNR$_{Y}$ & PCM$_{RR}$ & PointSSIM & PCQM & GraghSIM & PSNR$_\emph{p}$ & SSIM$_\emph{p}$ & MS-SSIM$_\emph{p}$ & VIFP$_\emph{p}$ \\\hline
   
      Bag               & 0.7018 & 0.5116 & 0.6025 & 0.5487  & 0.8124 & -0.5947 & 0.5750 & -0.8658 & 0.7600 & 0.7849 & 0.8476 & 0.8641 & 0.8771 & 0.8480\\
      Banana            & 0.7236 & 0.4086 & 0.5997 & 0.4086  & 0.7560 & -0.4797 & 0.1418 & -0.7145 & 0.5990 & 0.6337 & 0.7156 & 0.7812 & 0.7938 & 0.8724\\
      Biscuits          & 0.5258 & 0.5197 & 0.5633 & 0.5203  & 0.7812 & -0.3520 & 0.4850 & -0.7798 & 0.7490 & 0.5318 & 0.6953 & 0.7563 & 0.7775 & 0.8904\\
      Cake              & 0.4203 & 0.1327 & 0.2858 & 0.3577  & 0.5295 & -0.1209 & 0.1169 & -0.5832 & 0.4160 & 0.4848 & 0.6054 & 0.6096 & 0.6155 & 0.6743\\
      Cauliflower       & 0.4555 & 0.2914 & 0.3483 & 0.2914  & 0.6332 & -0.4199 & 0.1865 & -0.7057 & 0.6010 & 0.4847 & 0.6515 & 0.6068 & 0.6581 & 0.8578\\
      Flowerpot         & 0.7076 & 0.5271 & 0.6370 & 0.3816  & 0.6564 & -0.3224 & 0.3024 & -0.7030 & 0.6880 & 0.6779 & 0.8101 & 0.7963 & 0.8278 & 0.9368\\
      GlassesCase       & 0.6028 & 0.5132 & 0.5141 & 0.4370  & 0.7861 & -0.3922 & 0.4921 & -0.8214 & 0.7100 & 0.7277 & 0.7963 & 0.8025 & 0.8089 & 0.8077\\
      HoneydewMelon     & 0.4617 & 0.4337 & 0.4337 & 0.4337  & 0.7118 & -0.5654 & 0.5309 & -0.6539 & 0.7420 & 0.5291 & 0.7586 & 0.7543 & 0.8023 & 0.8989\\
      House             & 0.6391 & 0.3956 & 0.4312 & 0.3792  & 0.7972 & -0.4016 & 0.4155 & -0.7537 & 0.7410 & 0.6668 & 0.8311 & 0.8241 & 0.8257 & 0.8347\\
      Litchi            & 0.4291 & 0.3749 & 0.3472 & 0.3737  & 0.7201 & -0.4763 & 0.6147 & -0.7922 & 0.7260 & 0.6825 & 0.7685 & 0.8255 & 0.8545 & 0.9107\\
      Mushroom          & 0.6406 & 0.4860 & 0.5456 & 0.4580  & 0.8022 & -0.2575 & 0.4410 & -0.8033 & 0.7120 & 0.5700 & 0.7994 & 0.8296 & 0.8450 & 0.8697\\
      PenContainer      & 0.7782 & 0.5065 & 0.6688 & 0.5065  & 0.8132 & -0.5590 & 0.4916 & -0.8180 & 0.4080 & 0.8282 & 0.9183 & 0.9135 & 0.9153 & 0.9421\\
      Pineapple         & 0.4678 & 0.2923 & 0.3719 & 0.2923  & 0.7466 & -0.3341 & 0.4556 & -0.7578 & 0.0410 & 0.5280 & 0.7214 & 0.7134 & 0.7456 & 0.7817\\
      PingpongBat       & 0.7234 & 0.4191 & 0.6666 & 0.6320  & 0.8057 & -0.5132 & 0.5413 & -0.8600 & -0.2480 & 0.5595 & 0.6601 & 0.7189 & 0.7922 & 0.9096\\
      Pu'erTeaPot       & 0.3974 & 0.3688 & 0.3688 & 0.3688  & 0.8761 & -0.4041 & 0.4605 & -0.8400 & -0.1240 & 0.8084 & 0.8155 & 0.8658 & 0.8897 & 0.9201\\
      Pumpkin           & 0.5163 & 0.4919 & 0.4379 & 0.4919  & 0.6868 & -0.3401 & 0.3994 & -0.7462 & 0.2760 & 0.7250 & 0.8614 & 0.8606 & 0.8838 & 0.8976\\
      Ship              & 0.7676 & 0.3848 & 0.6505 & 0.5535  & 0.7918 & -0.3976 & 0.4134 & -0.7578 & -0.1990 & 0.7675 & 0.8201 & 0.8536 & 0.8791 & 0.9139\\
      Statue            & 0.8208 & 0.4298 & 0.7011 & 0.4648  & 0.7579 & -0.2564 & 0.3585 & -0.7860 & 0.2970 & 0.8364 & 0.9228 & 0.9184 & 0.9244 & 0.9623\\
      Stone             & 0.6140 & 0.5558 & 0.5161 & 0.5558  & 0.7882 & -0.3208 & 0.5102 & -0.8486 & 0.7850 & 0.7547 & 0.8282 & 0.8809 & 0.9233 & 0.8943\\
      ToolBox           & 0.4485 & 0.2923 & 0.2846 & 0.2846  & 0.9039 & -0.4919 & 0.4776 & -0.7662 & 0.8050 & 0.4777 & 0.5469 & 0.5723 & 0.7141 & 0.8532\\\hline
      Downsampling      & 0.4247 & 0.5408 & 0.3323 & 0.4437  & 0.6368 & -0.6681 & 0.9545 & -0.8863 & 0.9330 & 0.6783 & 0.8529 & 0.9375 & 0.9700 & 0.9767\\
      Gaussian noise    & 0.6867 & 0.6892 & 0.6867 & 0.6893  & 0.8706 & -0.7826 & 0.6743 & -0.9079 & 0.6200 & 0.8292 & 0.8213 & 0.8372 & 0.8467 & 0.9019\\
      G-PCC (T)         & 0.4018 & 0.3029 & 0.4050 & 0.3405  & 0.6322 & -0.3553 & 0.5843 & -0.8075 & 0.5680 & 0.3291 & 0.6065 & 0.6545 & 0.8105 & 0.8154\\
      V-PCC             & 0.1704 & 0.2156 & 0.2121 & 0.2866  & 0.3416 & -0.2805 & 0.3888 & -0.6563 & 0.4360 & 0.2903 & 0.3299 & 0.4397 & 0.7448 & 0.8419\\
      G-PCC (O)         &      0 &      0 & 0      & 0       & 0.8067 & -0.6084 & 0.8183 & -0.8935 & 0.5720 & 0.7730 & 0.8258 & 0.8774 & 0.8950 & 0.8943\\\hline
      \textbf{All}      & \textbf{0.4331} & \textbf{0.3425} & \textbf{0.3952} & \textbf{0.3412} & \textbf{0.6080} & \textbf{-0.3775} & \textbf{0.3436} & \textbf{-0.7486} & \textbf{0.4420} & \textbf{0.4989} & \textbf{0.6013} & \textbf{0.6701} & \textbf{0.7670} & \textbf{0.8504}\\
  \bottomrule

  \end{tabular}
  }
\end{table*}

\begin{table*}[t]
\centering
\caption{SRCC performance evaluation of the proposed model against existing models.}
\label{tab:srcc}
\scalebox{0.75}{
  \begin{tabular}{c|c c c c| c c c c c c c c c|c}

  \toprule
  \multirow{2}{*}{Subset}&
    \multicolumn{4}{c|}{Geometry distortion metric}&\multicolumn{9}{c|}{Geometry-plus-color distortion metric}& \multirow{2}{*}{IW-SSIM$_\emph{p}$}\\
    \cline{2-5} \cline{6-14} 
   
   & PSNR$_\emph{p2po,M}$ & PSNR$_\emph{p2po,H}$ & PSNR$_\emph{p2pl,M}$ & PSNR$_\emph{p2pl,H}$ & PSNR$_{Y}$ & PCM$_{RR}$ & PointSSIM & PCQM & GraghSIM & PSNR$_\emph{p}$ & SSIM$_\emph{p}$ & MS-SSIM$_\emph{p}$ & VIFP$_\emph{p}$ \\\hline
   
      Bag               & 0.6669 & 0.4363 & 0.5751 & 0.4365  & 0.8051 & -0.6069 & 0.3236 &  -0.8547 & 0.7320 & 0.7499 & 0.8438 & 0.8580 & 0.8725 & 0.8298\\
      Banana            & 0.6471 & 0.1933 & 0.5691 & 0.2033  & 0.6785 & -0.5287 & -0.0460 & -0.7686 & 0.5300 & 0.6759 & 0.7544 & 0.7968 & 0.7956 & 0.8627\\
      Biscuits          & 0.5252 & 0.3085 & 0.4160 & 0.3368  & 0.7719 & -0.4130 & 0.4865 & -0.7945 & 0.7250 & 0.5299 & 0.6757 & 0.7195 & 0.7380 & 0.8900\\
      Cake              & 0.3074 & 0.1724 & 0.1798 & 0.1796  & 0.5168 & -0.1503 & 0.0690 & -0.6043 & 0.4070 & 0.4365 & 0.5614 & 0.5602 & 0.5683 & 0.6598\\
      Cauliflower       & 0.3501 & 0.0918 & 0.2058 & 0.1653  & 0.5927 & -0.4718 & 0.2224 & -0.6971 & 0.5520 & 0.4305 & 0.5967 & 0.5730 & 0.5820 & 0.8125\\
      Flowerpot         & 0.6509 & 0.4348 & 0.5298 & 0.4515  & 0.6347 & -0.3058 & 0.3056 & -0.6984 & 0.6530 & 0.5963 & 0.7954 & 0.7776 & 0.8122 & 0.9211\\
      GlassesCase       & 0.5845 & 0.2020 & 0.4390 & 0.3238  & 0.7826 & -0.3883 & 0.2288 & -0.8137 & 0.6830 & 0.7624 & 0.8269 & 0.8205 & 0.8201 & 0.7845\\
      HoneydewMelon     & 0.4890 & 0.2768 & 0.3299 & 0.2300  & 0.6740 & -0.5742 & 0.4592 & -0.6439 & 0.7460 & 0.4512 & 0.7499 & 0.7207 & 0.7999 & 0.8954\\
      House             & 0.5866 & 0.3429 & 0.4483 & 0.3434  & 0.7826 & -0.4905 & 0.2968 & -0.7845 & 0.7500 & 0.7119 & 0.8312 & 0.8246 & 0.8267 & 0.8196\\
      Litchi            & 0.5109 & 0.3478 & 0.4291 & 0.3204  & 0.6496 & -0.4839 & 0.5026 & -0.7712 & 0.6840 & 0.6193 & 0.7231 & 0.8096 & 0.8556 & 0.8943\\
      Mushroom          & 0.6396 & 0.3486 & 0.5156 & 0.3105  & 0.6550 & -0.2556 & 0.4113 & -0.7819 & 0.6730 & 0.5863 & 0.7297 & 0.8535 & 0.8658 & 0.8528\\
      PenContainer      & 0.7720 & 0.2159 & 0.6688 & 0.3635  & 0.7963 & -0.6830 & 0.4059 & -0.8201 & 0.3760 & 0.8478 & 0.9372 & 0.9329 & 0.9334 & 0.9488\\
      Pineapple         & 0.3777 & 0.1376 & 0.2785 & 0.1831  & 0.7217 & -0.4011 & 0.3267 & -0.7862 & 0.0310 & 0.5334 & 0.7193 & 0.7105 & 0.7285 & 0.7584\\
      PingpongBat       & 0.5924 & 0.4958 & 0.4984 & 0.4357  & 0.7089 & -0.5526 & 0.4993 & -0.8224 & -0.2280 & 0.5420 & 0.6785 & 0.7236 & 0.7947 & 0.8945\\
      Pu'erTeaPot       & 0.6069 & -0.1173 & 0.4746 & -0.0384  & 0.8468 & -0.4308 & 0.3286 & -0.8528 & -0.1210 & 0.7432 & 0.7636 & 0.8414 & 0.8637 & 0.9170\\
      Pumpkin           & 0.4947 & 0.3092 & 0.3423 & 0.3068  & 0.6897 & -0.3241 & 0.2544 & -0.7802 & 0.1190 & 0.7347 & 0.8412 & 0.8497 & 0.8642 & 0.8831\\
      Ship              & 0.7464 & 0.3404 & 0.6267 & 0.5158  & 0.7734 & -0.4400 & 0.3578 & -0.7793 & -0.2050 & 0.7748 & 0.7847 & 0.8646 & 0.8829 & 0.9206\\
      Statue            & 0.8040 & 0.2450 & 0.6707 & 0.4487  & 0.6968 & -0.1811 & 0.3390 & -0.7570 & 0.1730 & 0.7947 & 0.9030 & 0.9118 & 0.9059 & 0.9561\\
      Stone             & 0.6219 & 0.3551 & 0.5129 & 0.3424  & 0.7115 & -0.3632 & 0.4924 & -0.8559 & 0.8040 & 0.6740 & 0.8303 & 0.8831 & 0.9203 & 0.8969\\
      ToolBox           & 0.3937 & 0.1972 & 0.2969 & 0.1884  & 0.8760 & -0.5239 & 0.3364 & -0.8473 & 0.8170 & 0.4720 & 0.5889 & 0.6268 & 0.7093 & 0.8307\\\hline
      Downsampling      & 0.4815 & 0.5356 & 0.3251 & 0.4879  & 0.6172 & -0.7407 & 0.8478 & -0.8760 & 0.7650 & 0.5399 & 0.8039 & 0.8876 & 0.9212 & 0.9270\\
      Gaussian noise    & 0.6155 & 0.6149 & 0.6194 & 0.6150  & 0.7895 & -0.7762 & 0.5931 & -0.8860 & 0.5760 & 0.6538 & 0.7509 & 0.7493 & 0.8067 & 0.8695\\
      G-PCC (T)         & 0.3451 & 0.2811 & 0.3568 & 0.3085  & 0.6247 & -0.3044 & 0.5669 & -0.8212 & 0.4780 & 0.1968 & 0.6144 & 0.6572 & 0.8153 & 0.8203\\
      V-PCC             & 0.1602 & 0.2051 & 0.1992 & 0.2370  & 0.3297 & -0.2966 & 0.3665 & -0.6431 & 0.2140 & 0.1998 & 0.3195 & 0.4213 & 0.7484 & 0.8458\\
      G-PCC (O)         &    NaN &    NaN & NaN    & NaN     & 0.8100 & -0.6468 & 0.8234 & -0.8944 & 0.4260 & 0.7809 & 0.8391 & 0.8770 & 0.8976 & 0.8981\\\hline
      \textbf{All}      & \textbf{0.4082} & \textbf{0.2578} & \textbf{0.3706} & \textbf{0.2883}  & \textbf{0.5849} & \textbf{-0.3603} & \textbf{0.3070} & \textbf{-0.7471} & \textbf{0.4360} & \textbf{0.4601} & \textbf{0.6138} & \textbf{0.6656} & \textbf{0.7689} & \textbf{0.8481}\\
  \bottomrule

  \end{tabular}
  }
\end{table*}
\subsection{Performance of Existing Objective PCQA Models}

Using the aforementioned database, we test the performance of 13 PCQA models, which are selected to cover a wide range of design methodologies. These models are chosen for two reasons. Firstly, geometry distortion metrics except the MPEG metrics are not included for assessing colored point clouds. Secondly, algorithms not publicly available are not included. The models include point-wise models: 1) point-to-point mean squared error-based PSNR ($\operatorname{PSNR}_{p2po,M}$)~\cite{tian2017evaluation,tian2017updated}, 2) point-to-point Hausdorff distance-based PSNR ($\operatorname{PSNR}_{p2po,H}$)~\cite{tian2017evaluation,tian2017updated}, 3) point-to-plane mean squared error-based PSNR ($\operatorname{PSNR}_{p2pl,M}$)~\cite{tian2017evaluation,tian2017updated}, 4) point-to-plane Hausdorff distance-based PSNR ($\operatorname{PSNR}_{p2pl,H}$)~\cite{tian2017evaluation,tian2017updated}, 5) point-to-point PSNR on color component ($\operatorname{PSNR}_{Y}$)~\cite{mekuria2017performance,mekuria2017design}, 6) PCM$_{RR}$~\cite{viola2020reduced}, 7) PointSSIM~\cite{alexiou2020towards}, 8) PCQM~\cite{meynet2020pcqm}, 9) GraghSIM~\cite{yang2020inferring}; and projection-based models: 10) projection-based PSNR ($\operatorname{PSNR}_p$)~\cite{torlig2018novel}, 11) projection-based structural similarity ($\operatorname{SSIM}_p$)~\cite{torlig2018novel,wang2004image}, 12) projection-based multi-scale structural similarity ($\operatorname{MS-SSIM}_p$)~\cite{torlig2018novel,wang2003multiscale}, and 13) projection-based pixel-domain visual information fidelity ($\operatorname{VIFP}_p$)~\cite{torlig2018novel,Sheikh2006image}. The implementation of all models are obtained from the original authors or their public websites.

We use PLCC, SRCC and RMSE between MOSs and model predictions as quantitative measures, and the test results are shown in Table~\ref{tab:plcc},~\ref{tab:srcc} and~\ref{tab:rmse}. We summarize the key observations as follows. First, it comes as no surprise that all geometry distortion models performs unfavorably to the geometry-plus-color PCQA model. Second, projection-based model, such as $\operatorname{VIFP}_p$, provide the most promising results so far. However, they often fall short in making a distinction of the perceptual importance between the background region and the regions corresponding to points in a 2D projection of a 3D point cloud. This suggests that more accurate objective PCQA models may be developed by removing the influence of background pixels and considering the perceptual importance of regions corresponding to the points. Third, even the best PCQA model only moderately correlates with human perception, leaving large space for improvement.

\begin{table*}[t]
\centering
\caption{RMSE performance evaluation of the proposed model against existing models.}
\label{tab:rmse}
\scalebox{0.75}{
  \begin{tabular}{c|c c c c| c c c c c c c c c|c}

  \toprule
  \multirow{2}{*}{Subset}&
    \multicolumn{4}{c|}{Geometry distortion metric}&\multicolumn{9}{c|}{Geometry-plus-color distortion metric}& \multirow{2}{*}{IW-SSIM$_\emph{p}$}\\
    \cline{2-5} \cline{6-14} 
   
   & PSNR$_\emph{p2po,M}$ & PSNR$_\emph{p2po,H}$ & PSNR$_\emph{p2pl,M}$ & PSNR$_\emph{p2pl,H}$ & PSNR$_{Y}$ & PCM$_{RR}$ & PointSSIM & PCQM & GraghSIM & PSNR$_\emph{p}$ & SSIM$_\emph{p}$ & MS-SSIM$_\emph{p}$ & VIFP$_\emph{p}$ \\\hline
   
      Bag               & 16.72 & 20.17 & 18.73 & 19.63  & 13.69 & 23.47 & 19.20 & 11.74 & 15.26 & 14.54 & 12.46 & 11.81 & 11.27 & 12.44\\
      Banana            & 14.98 & 19.97 & 17.36 & 19.81  & 14.20 & 21.70 & 21.48 & 15.18 & 17.38 & 16.79 & 15.16 & 13.54 & 13.20 & 10.61\\
      Biscuits          & 19.68 & 19.44 & 18.80 & 19.43  & 14.20 & 22.75 & 19.90 & 14.25 & 15.09 & 19.27 & 16.35 & 14.88 & 14.31 & 10.35\\
      Cake              & 20.55 & 22.31 & 21.57 & 21.02  & 19.10 & 22.51 & 22.36 & 18.29 & 20.47 & 19.69 & 17.92 & 17.84 & 17.74 & 16.62\\
      Cauliflower       & 19.95 & 21.44 & 21.01 & 21.44  & 17.35 & 22.41 & 22.02 & 15.88 & 17.91 & 19.60 & 17.00 & 17.82 & 16.88 & 11.52\\
      Flowerpot         & 16.81 & 20.24 & 18.33 & 21.98  & 17.94 & 23.78 & 22.67 & 16.92 & 17.26 & 17.48 & 13.95 & 14.39 & 13.34 & 8.322\\
      GlassesCase       & 18.06 & 19.68 & 19.41 & 20.36  & 13.99 & 22.63 & 19.72 & 12.91 & 15.94 & 15.52 & 13.69 & 13.50 & 13.31 & 13.34\\
      HoneydewMelon     & 20.99 & 21.36 & 21.34 & 21.33  & 16.62 & 23.66 & 20.05 & 17.90 & 15.85 & 20.08 & 15.42 & 15.53 & 14.12 & 10.37\\
      House             & 17.88 & 21.36 & 20.98 & 21.76  & 14.04 & 23.25 & 21.15 & 15.29 & 15.61 & 17.33 & 12.93 & 13.17 & 13.12 & 12.81\\
      Litchi            & 20.98 & 21.53 & 21.78 & 21.54  & 16.11 & 23.22 & 18.32 & 14.17 & 15.96 & 16.97 & 14.86 & 13.11 & 12.06 & 9.595\\
      Mushroom          & 17.02 & 19.37 & 18.58 & 19.71  & 13.23 & 22.17 & 19.89 & 13.20 & 15.56 & 18.21 & 13.32 & 12.38 & 11.85 & 10.94\\
      PenContainer      & 14.76 & 20.26 & 17.47 & 20.26  & 13.68 & 23.50 & 20.46 & 13.53 & 21.46 & 13.17 & 9.306 & 9.563 & 9.465 & 7.879\\
      Pineapple         & 18.04 & 19.52 & 18.95 & 19.53  & 13.58 & 20.41 & 18.17 & 13.32 & 20.20 & 17.34 & 14.14 & 14.30 & 13.61 & 12.73\\
      PingpongBat       & 15.69 & 20.64 & 16.94 & 17.61  & 13.46 & 22.73 & 19.11 & 11.60 & 22.73 & 18.84 & 17.07 & 15.80 & 13.87 & 9.442\\
      Pu'erTeaPot       & 21.75 & 22.15 & 22.05 & 22.11  & 11.43 & 23.71 & 21.04 & 12.86 & 23.71 & 13.96 & 13.72 & 11.86 & 10.82 & 9.288\\
      Pumpkin           & 18.63 & 18.94 & 19.56 & 18.94  & 15.81 & 21.75 & 19.94 & 14.48 & 20.91 & 14.98 & 11.05 & 11.08 & 10.18 & 9.591\\
      Ship              & 15.44 & 22.23 & 18.30 & 20.06  & 14.71 & 24.09 & 21.94 & 15.72 & 24.09 & 15.44 & 13.78 & 12.55 & 11.48 & 9.780\\
      Statue            & 13.24 & 20.93 & 16.53 & 20.53  & 15.12 & 23.18 & 21.64 & 14.33 & 22.19 & 12.71 & 8.932 & 9.170 & 8.841 & 6.309\\
      Stone             & 16.85 & 18.48 & 18.28 & 17.76  & 13.14 & 21.35 & 18.36 & 11.29 & 13.23 & 14.00 & 11.96 & 10.10 & 8.201 & 9.551\\
      ToolBox           & 18.65 & 19.95 & 20.01 & 20.00  & 8.923 & 20.86 & 18.33 & 13.41 & 12.37 & 18.33 & 17.47 & 17.11 & 14.61 & 10.88\\\hline
      Downsampling      & 26.93 & 25.02 & 28.07 & 26.66  & 22.93 & 29.74 & 8.870 & 13.79 & 10.68 & 21.86 & 15.53 & 10.35 & 7.229 & 6.384\\
      Gaussian noise    & 14.15 & 14.11 & 14.15 & 14.10  & 9.576 & 19.47 & 14.38 & 8.162 & 15.28 & 10.88 & 11.11 & 10.65 & 10.36 & 8.408\\
      G-PCC (T)         & 22.59 & 23.51 & 22.56 & 23.20  & 19.11 & 24.67 & 20.02 & 14.64 & 20.30 & 23.30 & 19.61 & 18.65 & 14.45 & 14.28\\
      V-PCC             & 16.76 & 16.61 & 16.62 & 16.29  & 15.99 & 17.01 & 15.67 & 12.84 & 15.30 & 16.28 & 16.06 & 15.28 & 11.35 & 9.179\\
      G-PCC (O)         & 21.36 & 21.36 & 21.36 & 21.36  & 12.62 & 21.36 & 12.28 & 9.593 & 17.53 & 13.55 & 12.05 & 10.25 & 9.531 & 9.559\\\hline
      \textbf{All}      & \textbf{20.66} & \textbf{21.54} & \textbf{21.06} & \textbf{21.55} & \textbf{18.20} & \textbf{22.92} & \textbf{21.53} & \textbf{15.20} & \textbf{20.56} & \textbf{19.87} & \textbf{18.32} & \textbf{17.02} & \textbf{14.71} & \textbf{12.06}\\
  \bottomrule

  \end{tabular}
  }
\end{table*}

\section{Objective Quality Assessment}\label{sec:objective}

\subsection{Proposed PCQA Model}

A point cloud can be omni-directionally inspected from a view-sphere at a given distance, while it is both cumbersome and unnecessary to use a large number of viewpoints when acquiring its 2D snapshots. Icosphere, a unit geodesic sphere created by subdividing a regular icosahedron with normalized vertices, are employed to generate viewpoints due to the uniform distribution of vertices~\cite{alexiou2019exploiting,lavoue2015efficiency}. The number of vertices that can be generated is
\begin{equation} \label{eq:number}
N_v=12+10\left(4^l-1\right),
\end{equation}
where $l$ represents the subdivision level. For any point in a 3D point cloud, let $\textbf{p}=(\textbf{g}\ \textbf{c})$ be a 6 dimensional row vector where $\textbf{g}$ and $\textbf{c}$ contain its 3D coordinates $\left(g_x\ g_y\ g_z\right)$ and attributed color information $\left(c_r\ c_g\ c_b\right)$, respectively. We use a series of transformations to obtain the projected images.

Firstly, we translate a point cloud to align its geometric center to origin $\left(0\ 0\ 0\right)$. Specifically, for each $\textbf{p}$
\begin{equation} \label{eq:translation}
\textbf{g}_{t}=\textbf{g}-\textbf{t}_{r},
\end{equation}
where $\textbf{g}_{t}$ represents the 3D coordinates after translation, and $\textbf{t}_{r}$ represents the translation vector equalling the geometric center coordinates of its corresponding reference point cloud. The reason $\textbf{t}_{r}$ is used instead of the geometric center of a distorted point cloud is that geometric distortion may cause changes of the upper and lower bounds of 3D coordinates, leading to misalignment of the projected images.

Secondly, we rotate the point cloud to obtain a number of viewpoints. More specifically, let $\textbf{n}_{v}$, a 3 dimensional row vector, be the unit normal and $\textbf{n}_{z}$ be $\left(0\ 0\ 1\right)$, then the rotation vector $\left(\textbf{r}\ \theta\right)$ can be calculated as 
\begin{equation} \label{eq:r}
\textbf{r}=\frac{\textbf{n}_{v}\times \textbf{n}_{z}}{\|\textbf{n}_{v}\times \textbf{n}_{z}\|}
\end{equation}
and
\begin{equation} \label{eq:theta}
\theta=\arccos{\left(\textbf{n}_{v}\cdot \textbf{n}_{z}\right)}, 
\end{equation}
where $\| \cdot \|$ denotes the $l^2$ norm of a vector, $\textbf{r}$ is the rotation axis, and $\theta$ is the axial angle. The rotation matrix $\textbf{R}$ is obtained using $\left(\textbf{r}\ \theta\right)$. Then we use $\textbf{R}$ to calculate $\textbf{g}_{r}$, the 3D coordinates after rotation, for each $\textbf{p}$,

\begin{equation} \label{eq:rotation}
\textbf{g}_{r}=\textbf{g}_{t}\textbf{R}. 
\end{equation}

Thirdly, scaling transformation is applied to make 2D snapshots of all reference point clouds approximately watertight meanwhile keeping details as much as possible. For each $\textbf{p}$, this operation can be expressed as
\begin{equation} \label{eq:scaling}
\textbf{g}_{s}=s\cdot\textbf{g}_{r}, 
\end{equation}
where $\textbf{g}_{s}$ represents 3D coordinates after scaling and $s$ is a scaling factor. Since the values of the coordinates are rounded to integer numbers, for which the maximum rounding error is bounded by half of the pixel spacing, the default value of $s$ is set to 1/2. Empirically, we also find this leads to the best performance.

Fourthly, we use orthogonal projection~\cite{torlig2018novel} and rasterization to obtain a projected image. For each \textbf{p}, the projected coordinates are given by
\begin{equation} \label{eq:projection}
\widetilde{\textbf{g}} =\textbf{g}_{s}\textbf{P},
\end{equation}
where
\begin{equation} \label{eq:projectionmatrix}
\textbf{P}=
\left[\begin{array}{ccc}   
1 & 0 & 0\\
0 & 1 & 0\\
0 & 0 & 0\\
\end{array}\right]
\end{equation}
represents orthogonal projection matrix. If there are multiple points occupying the same location $\left(\widetilde{g}_x\ \widetilde{g}_y\right)$, the point with the largest value of $g_z$ will be maintained. To implement rasterization, we put \textbf{c} of all maintained points into their corresponding projected positions on a projection plane filled with \textbf{c} of $\left(127\ 127\ 127\right)$, and we obtain a projected image denoted as $\textbf{I}$. 

Finally, we perform the above operations on each distorted point cloud $PC_{dis}$ and its reference $PC_{ref}$ to obtain $\textbf{I}_{dis}\left(n\right)$ and $\textbf{I}_{ref}\left(n\right)$ at the $n$th viewpoint of icosphere, respectively.

\begin{figure*}
    \centering
    \subfloat[]{\includegraphics[width=0.25\textwidth]{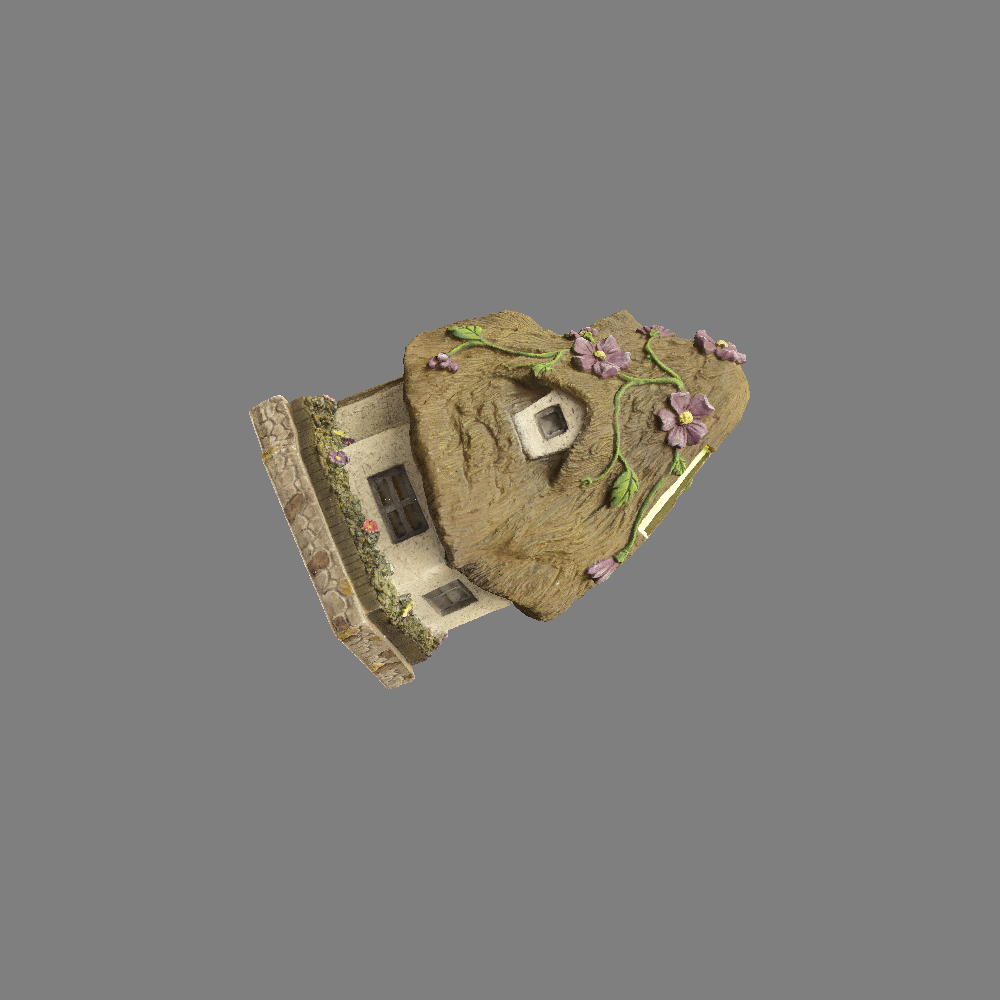}}\hskip0.1em
    \subfloat[]{\includegraphics[width=0.25\textwidth]{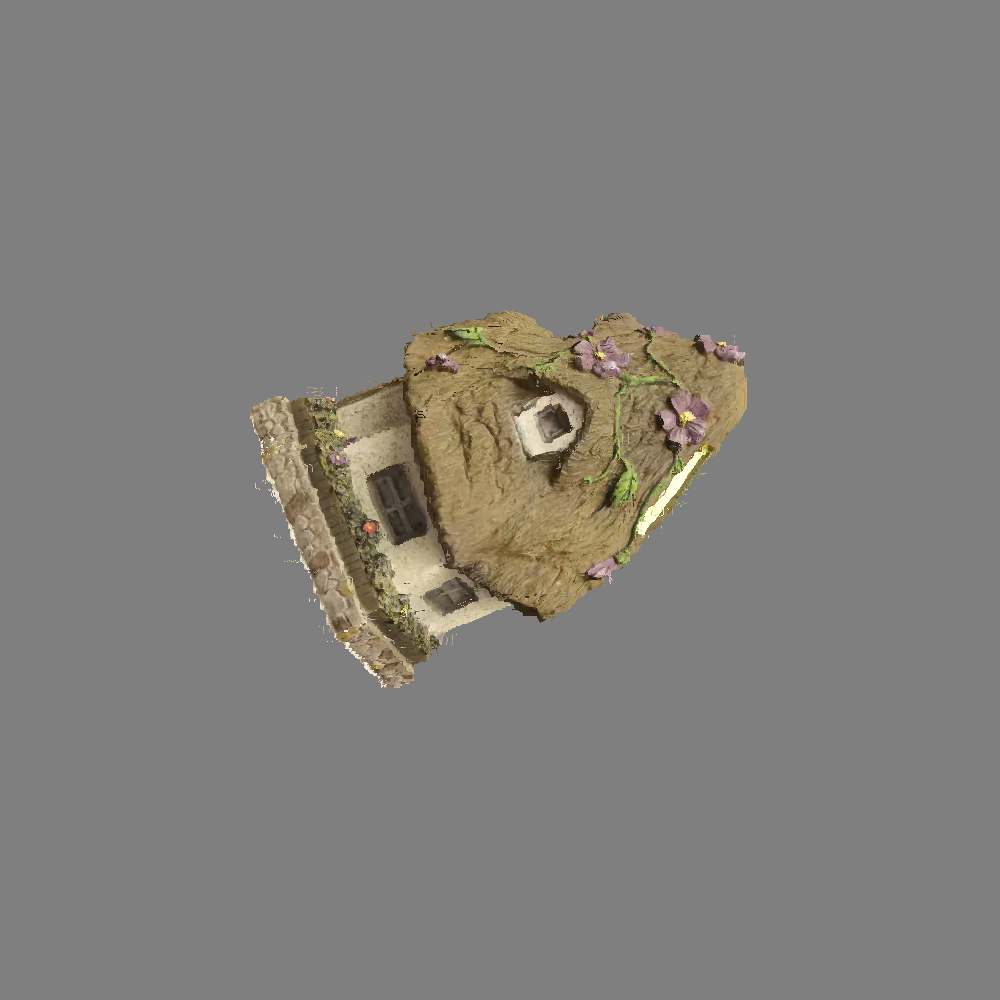}}\hskip0.1em
    \subfloat[]{\includegraphics[width=0.25\textwidth]{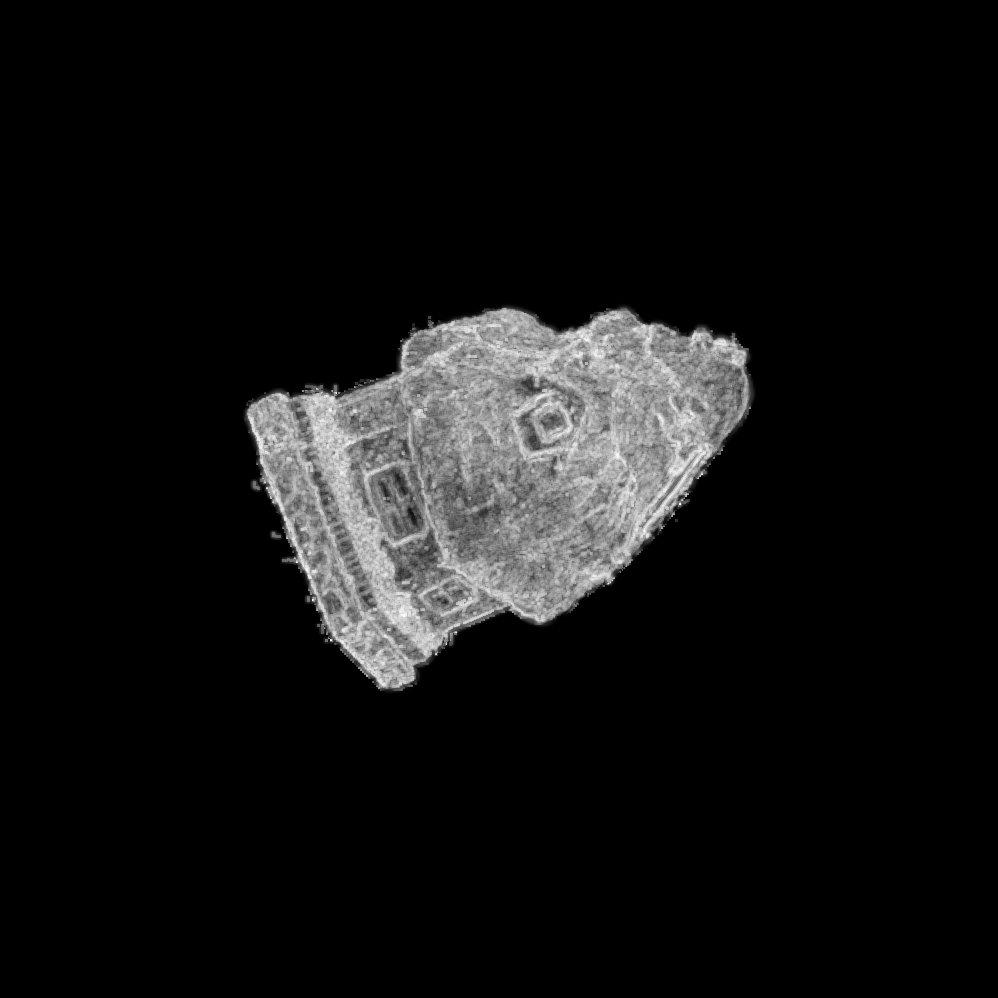}}\hskip0.1em
    \subfloat[]{\includegraphics[width=0.125\textwidth]{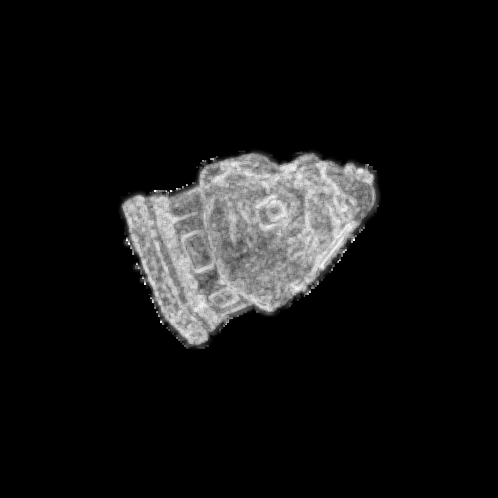}}\hskip0.1em
    \subfloat[]{\includegraphics[width=0.0625\textwidth]{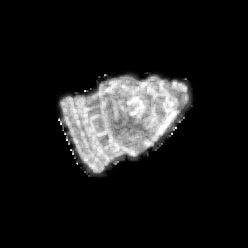}}\hskip0.1em
    \subfloat[]{\includegraphics[width=0.03125\textwidth]{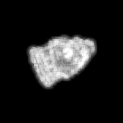}}\hskip0.1em
    \caption{Local information content maps. (a), (b) Snapshots of ``House'' and one of its distorted versions. (c), (d), (e), (f) Information content maps computed at four scales.}
    \label{fig:IWmaps}
\end{figure*}

 $\textbf{I}_{dis}\left(n\right)$ and $\textbf{I}_{ref}\left(n\right)$ have identical background pixels, and thus, the similarity between $\textbf{I}_{dis}\left(n\right)$ and $\textbf{I}_{ref}\left(n\right)$ is larger than that of $PC_{dis}$ and $PC_{ref}$. Consequently, for better distinction, it is useful to remove the influence of background pixels. It is also worth noting that in a 2D projection of a 3D point cloud, the perceptual importance of different regions changes significantly over space. In particular, the background region contains no information about the point clouds, and the importance of the regions corresponding to the points in the cloud also varies. The principle of information content weighted pooling~\cite{wang2010information} provides an excellent framework to account for such variations in importance, as exemplified by Fig.~\ref{fig:IWmaps}. Therefore, we propose to assess the perceptual quality of $PC_{dis}$, by
\begin{equation} \label{eq:projectionIWSSIM}
Q\left(PC_{dis}\right) =\sum_{n=1}^{N_v}{\operatorname{IW-SSIM}\left(\textbf{I}_{ref}\left(\emph{n}\right),\textbf{I}_{dis}\left(\emph{n}\right)\right)},
\end{equation}
where the IW-SSIM evaluations~\cite{wang2010information} between all pairs of $\textbf{I}_{ref}\left(n\right)$ and $\textbf{I}_{dis}\left(n\right)$ are averaged to an overall quality measure of $PC_{dis}$, and we name the proposed method IW-SSIM$_\emph{p}$, where all default parameters of IW-SSIM~\cite{wang2010information} in still image quality assessment are inherited.
\subsection{Validation and Discussion}
\begin{table*}[t]
\centering
\caption{STATISTICAL SIGNIFICANCE COMPARISON MATRIX BASED ON QUALITY PREDICTION RESIDUALS. A SYMBOL ``1'' MEANS THAT THE PERFORMANCE OF THE ROW MODEL IS STATISTICALLY BETTER THAN THAT OF THE COLUMN MODEL, A SYMBOL ``0'' MEANS THAT THE ROW MODEL IS STATISTICALLY WORSE, AND A SYMBOL ``-'' MEANS THAT THE ROW AND COLUMN MODELS ARE STATISTICALLY INDISTINGUISHABLE.}
\label{tab:statistical}
\scalebox{0.75}{
  \begin{tabular}{c|c c c c c c c c c c c c c c}

  \toprule
   
   & PSNR$_\emph{p2po,M}$ & PSNR$_\emph{p2po,H}$ & PSNR$_\emph{p2pl,M}$ & PSNR$_\emph{p2pl,H}$ & PSNR$_\emph{Y}$ & PCM$_{RR}$ & PointSSIM & PCQM & GraghSIM & PSNR$_\emph{p}$ & SSIM$_\emph{p}$ & MS-SSIM$_\emph{p}$ & VIFP$_\emph{p}$ & IW-SSIM$_\emph{p}$\\\hline
   PSNR$_\emph{p2po,M}$ &-&-&-&-&0&-&-&0&-&-&0&0&0&0\\
   PSNR$_\emph{p2po,H}$ &-&-&-&-&0&-&-&0&-&0&0&0&0&0\\
   PSNR$_\emph{p2pl,M}$ &-&-&-&-&0&-&-&0&-&0&0&0&0&0\\
   PSNR$_\emph{p2pl,H}$ &-&-&-&-&0&-&-&0&-&0&0&0&0&0\\
   PSNR$_\emph{Y}$      &1&1&1&1&-&1&1&0&1&-&-&0&0&0\\
   PCM$_{RR}$           &-&-&-&-&0&-&-&0&-&0&0&0&0&0\\
   PointSSIM            &-&-&-&-&0&-&-&0&-&0&0&0&0&0\\
   PCQM                 &1&1&1&1&1&1&1&-&1&1&1&1&-&0\\
   GraphSIM             &-&-&-&-&0&-&-&0&-&-&0&0&0&0\\
   PSNR$_\emph{p}$      &-&1&1&1&-&1&1&0&-&-&-&0&0&0\\
   SSIM$_\emph{p}$      &1&1&1&1&-&1&1&0&1&-&-&0&0&0\\
   MS-SSIM$_\emph{p}$   &1&1&1&1&1&1&1&0&1&1&1&-&0&0\\
   VIFP$_\emph{p}$      &1&1&1&1&1&1&1&-&1&1&1&1&-&0\\
   IW-SSIM$_\emph{p}$   &1&1&1&1&1&1&1&1&1&1&1&1&1&-\\
   
  \bottomrule

  \end{tabular}
  }
\end{table*}
We validate the proposed IW-SSIM$_\emph{p}$ model using the WPC database presented in Section~\ref{sec:subjective} and compare its performance against existing objective PCQA models. Note that IW-SSIM$_\emph{p}$ does not involve a training process and is independent of any existing PCQA databases including the WPC database. Tables~\ref{tab:plcc}, \ref{tab:srcc} and~\ref{tab:rmse} summarize the PLCC, SRCC and RMSE evaluation results. We find that evaluation results of IW-SSIM$_\emph{p}$ when $Nv=12,42,162$ are very close to each other, while the computational complexity is proportional to $Nv$. Therefore we use $Nv=12$ in all results reported here. It can be seen that the proposed model delivers the best performance in predicting subjective quality of 3D point cloud not only on the whole database but also on almost every subset. In addition, its PLCC and SRCC performance is at the same level as compared to an average human subject as in Fig.\ref{fig:plccsrcc}. 


To ascertain that the improvement of the proposed model is statistically significant, we carried out a statistical significance analysis by following the approach introduced in\cite{sheikh2006statistical}. First, a nonlinear regression function is applied to map the objective quality scores to predict the subjective scores. We observe that the prediction residuals all have zero-mean, and thus the model with lower variance is generally considered better than the one with higher variance. We conduct a hypothesis testing using F-statistics. Since the number of samples exceeds 50, the Gaussian assumption of the residuals approximately hold based on the central limit theorem\cite{montgomery2014applied}. The test statistic is the ratio of variances. The null hypothesis is that the prediction residuals from one quality model come from the same distribution and are statistically indistinguishable (with 95\% confidence) from the residuals from another model. We compare every possible pairs of objective models. The results are summarized in Table~\ref{tab:statistical}, where a symbol ``1'' means that the row model performs significantly better than the column model, a symbol ``0'' means the opposite, and a symbol ``-'' indicates that the row and column models are statistically indistinguishable. 

There are several useful findings from the statistical significance analysis. First, existing geometry distortion metrics are statistically indistinguishable from each other. Second, most geometry-plus-color distortion metrics are statistically better than geometry distortion metrics. Third, the proposed IW-SSIM$_\emph{p}$ model is statistically better than all existing models.

Finally, to investigate the generalization potential of the proposed model, we performed a cross-database validation on popular databases and compared our metric with the well-known state-of-the-art GraphSIM~\cite{yang2020inferring}, PointSSIM~\cite{alexiou2020towards}, PCQM~\cite{meynet2020pcqm} and PCM$_{RR}$~\cite{viola2020reduced}. Table \ref{tab:plcc_cross}, \ref{tab:srcc_cross} and \ref{tab:rmse_cross} depict the results of cross-database validation. We can draw the following conclusions. First, IW-SSIM$_\emph{p}$ performs well on all databases except the IRPC database. Notice that it seems most of the PCQA models performs not very well on this database. Second, PCQM is the best metric for SJTU-PCQA database, meanwhile, GraghSIM has the best performance on the IRPC and M-PCCD database. Third, GraphSIM is a competitive model for IW-SSIM$_\emph{p}$, however, its time complexity is too high in the actual experiment.
\begin{table}[t]
\centering
\caption{PLCC performance evaluation for cross-database validation.}
\label{tab:plcc_cross}
\scalebox{0.8}{
  \begin{tabular}{c|c c c c c}

  \toprule
   
   & GraghSIM & PointSSIM & PCQM & PCM$_{RR}$ & IW-SSIM$_\emph{p}$ \\\hline
   
   SJTU-PCQA~\cite{yang2020predicting} & 0.5910 & 0.7503 & \textbf{-0.8565} & -0.5129 & 0.7949\\
       IRPC~\cite{javaheri2020point}      & \textbf{0.8603} & 0.5939 & -0.1850 & -0.0540 & 0.0911\\
   ICIP2020~\cite{perry2020quality}  & 0.8601 & 0.6758 & -0.2634 & -0.7233 & \textbf{0.9097}\\
   M-PCCD~\cite{alexiou2019comprehensive}    & \textbf{0.9428} & 0.8519 & -0.6070 & -0.5535 & 0.7172\\
   WPC       & 0.4420 & 0.3436 & -0.7486 & -0.3775 & \textbf{0.8504}\\
  \bottomrule

  \end{tabular}
  }
\end{table}
\begin{table}[t]
\centering
\caption{SRCC performance evaluation for cross-database validation.}
\label{tab:srcc_cross}
\scalebox{0.8}{
  \begin{tabular}{c|c c c c c}

  \toprule
   
   & GraghSIM & PointSSIM & PCQM & PCM$_{RR}$ & IW-SSIM$_\emph{p}$ \\\hline
   
   SJTU-PCQA~\cite{yang2020predicting} & 0.5710 & 0.7346 & \textbf{-0.8439} & -0.5366 & 0.7833\\
   IRPC~\cite{javaheri2020point}      & \textbf{0.7469} & 0.5054 & -0.4170 & -0.2345 & 0.1339\\
   ICIP2020~\cite{perry2020quality}  & 0.8449 & 0.5638 & -0.4108 & -0.8868 & \textbf{0.8968}\\
   M-PCCD~\cite{alexiou2019comprehensive}    & \textbf{0.9535} & 0.8328 & -0.9155 & -0.8885 & 0.7487\\
   WPC       & 0.4360 & 0.3070 & -0.7471 & -0.3603 & \textbf{0.8481}\\
  \bottomrule

  \end{tabular}
  }
\end{table}
\begin{table}[t]
\centering
\caption{RMSE performance evaluation for cross-database validation.}
\label{tab:rmse_cross}
\scalebox{0.8}{
  \begin{tabular}{c|c c c c c}

  \toprule
   
   & GraghSIM & PointSSIM & PCQM & PCM$_{RR}$ & IW-SSIM$_\emph{p}$ \\\hline
   
   SJTU-PCQA~\cite{yang2020predicting} (10) & 1.8910 & 1.5499 & \textbf{1.2169} & 2.3442 & 1.4224\\
   IRPC~\cite{javaheri2020point} (5)      & \textbf{0.5207} & 0.7944 & 0.9703 & 0.9874 & 0.9833\\
   ICIP2020~\cite{perry2020quality} (5)  & 0.5794 & 0.8373 & 1.1360 & 1.1360 & \textbf{0.4718}\\
   M-PCCD~\cite{alexiou2019comprehensive} (5)   & \textbf{0.4535} & 0.7124 & 1.3603 & 1.3603 & 0.9480\\
   WPC (100)      & 20.5640 & 21.5273 & 15.1996 & 22.9234 & \textbf{12.0620}\\
  \bottomrule

  \end{tabular}
  }
\end{table}
\section{Conclusion}\label{sec:conclusion}
In this work, we tackle the problem of 3D point cloud quality assessment. Our major contributions are fourfold. First, we construct 20 high quality, realistic and omni-directional dense point clouds with diverse geometric and textural complexity, which are voxelized with an average number of 1.35M points and a standard deviation of 656K, respectively. These point clouds not only can be used for PCQA, but also are useful to other fields of point cloud processing. Second, we construct so far the largest point cloud database of diverse content and distortion variations and conduct a lab-controlled subjective user study. The proposed WPC database contains 740 distorted point clouds with MOSs approximately evenly distributed from poor to excellent perceived quality levels. Third, we conduct a comprehensive evaluation on existing objective PCQA models and find that state-of-the-art PCQA models do not provide reliable predictions of perceived quality. Fourth, we propose a projection-based IW-SSIM$_\emph{p}$ model that significantly outperforms existing objective PCQA methods.

\ifCLASSOPTIONcompsoc
  \section*{Acknowledgments}
\else
  \section*{Acknowledgment}
\fi

This work was supported in part by Natural Sciences and Engineering Research Council of Canada, in part by the National Natural Science Foundation of China under Grants (61772294, 62172259), in part by Shandong Provincial Natural Science Foundation of China under Grant ZR2018PF002, in part by the open project program of state key laboratory of virtual reality technology and systems, Beihang University, under Grant VRLAB2021A01, in part by the Joint funding for smart computing of Shandong Natural Science Foundation of China under Grant ZR2019LZH002, and jointly completed by the State Key Laboratory of High Performance Server and Storage Technology, Inspur Group, Jinan, China.

\ifCLASSOPTIONcaptionsoff
  \newpage
\fi



%
\bibliographystyle{IEEEtran}
\bibliography{ref}

\end{document}